\documentclass[a4paper,11pt]{article}
\pdfoutput=1
\usepackage{jheppub} 
\usepackage[]{epsfig}
\usepackage{amsfonts}
\usepackage{amsmath}
\usepackage{color}
\usepackage{float}
\usepackage{slashed}
\usepackage{subfigure}
\usepackage{pstricks}
\usepackage{graphicx}
\usepackage{graphics}
\def\ba{\begin{eqnarray}}
\def\ea{\end{eqnarray}}
\def\be{\begin{equation}}
\def\ee{\end{equation}}

\def\uz{\underline z}
\def\ut{\underline t}

\def\ux{\underline x}
\def\uy{\underline y}
\newcommand{\bea}{\begin{eqnarray}} 
\newcommand{\eea}{\end{eqnarray}}

\title{\boldmath Pomeranchuk instabilities in holographic metals}
\author{Gast\'on Giordano,}
\author{Nicol\'as Grandi,}
\author{Adri\'an Lugo,}
\affiliation{Instituto de F\'\i sica La Plata IFLP-CONICET \&  Departamento de F\'isica, FCE-UNLP\\  C.C. 67, 1900 La Plata, Argentina}
\emailAdd{gaston.giordano@fisica.unlp.edu.ar}
\emailAdd{grandi@fisica.unlp.edu.ar}
\emailAdd{lugo@fisica.unlp.edu.ar}
\abstract{We develop a method to detect instabilities leading to nematic phases in strongly coupled metallic systems. We do so by adapting the well-known Pomeranchuk technique to a weakly coupled system of fermions in a curved asymptotically AdS bulk. The resulting unstable modes are interpreted as corresponding to instabilities on the dual strongly coupled holographic metal. We apply our technique to a relativistic $3+1$-dimensional bulk with generic quartic fermionic couplings, and explore the phase diagram at zero temperature for finite values of the fermion mass and chemical potential, varying the couplings. We find a wide region of parameters where the system is stable, which is simply connected and localized around the origin of coupling space. }
\begin{document}
\maketitle
\newpage
\section{Introduction}
\label{sec:introduction}
Strongly correlated electron systems have been at the heart of most recent research in condensed matter theory \cite{sachdev_2000}. The underlying strongly coupled dynamics is thought to be responsible of the richness of the phase diagram of high $T_c$ superconductors. It contains normal metallic regions, as well as an exotic phase known as ``strange metal'' or ``non-Fermi liquid'' \cite{RevModPhys.73.797}  which is believed to be described by strongly coupled fermionic degrees of freedom. There are also regions in which rotational symmetry is broken, the so-called ``nematic'' phases, as well as inhomogeneous ``smectic'' and ``chessboard'' phases.

The transition from an isotropic Fermi liquid to a nematic phase is believed to be driven by a Pomeranchuk instability \cite{pomeranchuk}. Such instability arises when an excitation of the ground state of the Fermi liquid results in a net decrease of the total energy. In Landau theory, such perturbation is represented by a deformation of the Fermi surface. By decomposing the deformation onto an orthonormal basis,  
Pomeranchuk obtained a set of conditions under which the Fermi liquid is stable. This method can be generalized to lattice systems or anisotropic Fermi surfaces \cite{PhysRevB.78.115104, PhysRevB.80.075108} 
and to finite temperature and magnetic field \cite{uno, dos, RodriguezPonte2013}, 
and it relies on the weakness of the quasi-particle coupling, or in other words on the validity of the Landau formula. This implies that from the strange metal perspective, since the dynamics is strongly coupled, the detection of fermionic instabilities becomes more difficult.

The holographic description \cite{AHARONY2000183} of a hom\-ogen\-eous fermionic phase has been shown to account for some of the interesting properties of the strange metal \cite{Lee:2008xf, Liu:2009dm,Faulkner:2009wj,Faulkner:2010da, Hartnoll:2010gu,Hartnoll:2011dm}. In particular, the resulting spectral function is compatible with a Fermi surface with or without long-lived quasi-particles. Based on that, in this paper, we analyze Pomeranchuk instabilities of the strange metal phase from the holographic perspective. We use a holographic background in which we propagate a Dirac spinor. Being weakly coupled, such spinor can be described by Landau's theory, and its stability under Fermi surface deformations can be studied. This accounts for a description of the anisotropic instabilities of the dual strange metal. 

The generalization of Pomeranchuk method to arbitrary curved spaces is not possible, since it relies on a momentum space representation of the fermionic system. Nevertheless, the special kind of ``planar'' bulk spacetimes used in holography have the additional feature of a translational symmetry in the spatial directions spanning the boundary. This allows for a labeling of the bulk states by a momentum index $\vec k$, complemented by an additional index $m$ labeling the oscillation mode in the holographic direction. Then the $D+1$ dimensional curved space fermionic system can be interpreted as a $(D-1)+1$ dimensional flat space system with multiple fermion species labeled by $m$. Pomeranchuk method can then be straightforwardly applied to such multi-fermion flat space weakly coupled system.

In what follows, we sketch the necessary steps needed to go from a $3+1$ dimensional action for spinors in AdS spacetime to a $2+1$ dimensional Hamiltonian for a multi-fermion system in flat space. Then we construct the corresponding Landau theory, and apply the Pomeranchuk method to it. To improve readability, only the relevant steps of the calculations are shown in the bulk of the paper, leaving the details to the appendices \footnote{Notice that Pomeranchuk instabilities were studied in the holographic context in \cite{Edalati:2012eh} from a different perspective, focusing in the distortions of spectral functions. See also \cite{Liu:2014mva} for a related issue.}.

\newpage
\section{Holographic setup}
\label{sec:holographic.setup}
We consider a holographic background consisting of a metric and an electromagnetic field with the generic ``planar'' form
\bea
\label{eq:background}
G &=& L^2\left(
-f \, dt^2+g\, dz^2 +\frac{dx^2+dy^2}{z^2}
\right)\,,\qquad\qquad
A=h\,dt\,.
\eea
This is a solution of the Einstein-Maxwell equations with a negative cosmological constant,
as well as possible additional matter contributions to the energy-momentum tensor. The metric components $f(z)$ and $g(z)$ are functions of the holographic coordinate $z$, and the geometry asymptotes AdS spacetime as $z$ goes to zero, provided $f(z)\sim g(z)\sim 1/z^2$. The electric potential $h(z)$ is also a function of $z$, and approaches a constant $h(z)\sim \mu$  at the boundary, which is identified with the chemical potential on the holographic theory.

Specific backgrounds with the form \eqref{eq:background} are the Reisner-Nordstrom AdS black-hole \cite{Lee:2008xf, Liu:2009dm, Faulkner:2010da, Faulkner:2009wj}, the holographic superconductor \cite{Hartnoll_2008}, and the electron star at zero \cite{Hartnoll:2010gu, Hartnoll:2011dm, Cubrovic:2011xm} and finite \cite{Puletti:2010de, Hartnoll:2010ik} temperatures. 
Although we keep our formalism as general as possible, in order to get concrete results in Section \ref{sec:application.example} we apply it to the  background of reference \cite{Hartnoll:2010gu}.

\bigskip

In the background defined above we propagate a spinorial perturbation $\Psi$, whose dynamics is dictated by the free Dirac action
\be
\label{eq:dirac.action}
S_{\sf free}= -\int d^4x\,\sqrt{|G|}\; 
\bar \Psi \left(\slash\!\!\!\!{\cal D}   
- {\sf m}
\right)\Psi \,,
\ee
where $\slash\!\!\!\!{\cal D}$ stands for the covariant derivative containing both curved spacetime and gauge contributions, contracted with the curved space Dirac matrices. Since the energy-mo\-men\-tum tensor and the electric current are quadratic in the spinor perturbation, to linear order in $\Psi$ we can work in the probe limit in which the background \eqref{eq:background} is not perturbed.

The general solution to \eqref{eq:dirac.action} can be decomposed as 
\bea
\label{eq:fermion.decomposition}
\Psi(t,\vec x,z)&=& \frac{z}{f(z)^\frac{1}{4}}\,\sum_{\alpha m\vec k} \frac1{{\cal N}_{\alpha m\vec k}} \;\,c_{\alpha m\vec k}(t)
\;\,e^{i \vec k\cdot\vec x}\; \psi_{\alpha m \vec k}(z)\,,
\eea
in terms of time dependent coefficients $c_{\alpha m\vec k}(t)$ and $z$-dependent spinors  $\psi_{\alpha m \vec k}(z)$. Here the label $\alpha=1, 2$ is a spin index, and $\vec k\in \mathbb{R}^2$ represents the momentum in the $xy$ plane, while $m\in\mathbb{N}$ characterizes the number of oscillations  in the $z$ direction. Finally ${\cal N}_{\alpha m\vec k}$ is a normalization constant and the factor $z/f^{1/4}(z)$  was introduced for calculational convenience.
The spinors $\psi_{\alpha m\vec k}(z)$ are written in terms of the solution to an ordinary differential equation in the variable $z$, with a Schr\"odinger-like form. When physically meaningful boundary conditions are imposed at the extremes of the $z$ range, a (possibly complex) quantized dispersion relation is obtained $\omega_m(k)$, with which we can write $c_{\alpha m\vec k}(t)\propto\exp(-i\omega_m(k)t)$.

Expression \eqref{eq:fermion.decomposition} allows us to quantize the system by promoting the coef\-ficien\-ts $c_{\alpha m\vec k}(t)$  to operators $c_{\alpha m\vec k}$ and $c^\dagger_{\alpha m\vec k}$ satisfying fermionic anticommutation relations. Then $c_{\alpha m \vec k}^\dagger$ creates a fermionic perturbation with momentum $\vec k$, spin $\alpha$ and mode index $m$, and $c_{\alpha m \vec k}$ annihilates it.

\bigskip

The steps of the derivation leading to decomposition \eqref{eq:fermion.decomposition} as well as the explicit form of the components of the spinor $\psi_{\alpha m k}(z)$ are reviewed in detail in Appendix \ref{app:free.fermionic.states}.

\section{Interactions}
\label{{sec:interactions}}
Now we introduce interactions among the spinorial perturbations, which represent $1/N$ corrections to the holographic description. We do so by supplementing the action with the additional term
\be 
\label{eq:interaction.action}
S_{\sf int}= \int d^4x\,\sqrt{|G|}\; 
 T^{\bar\sigma\bar\sigma'}_{\sigma\sigma'}\;  \bar\Psi_{\bar\sigma}\;\bar\Psi_{\bar\sigma'}\;\Psi^{\sigma}\;\Psi^{\sigma'}\,,
\ee
where we made explicit the spinor indices $\sigma\in\{1,2,3,4\}$, and the Lorentz invariant tensor $T^{\bar\sigma\bar\sigma'}_{\sigma\sigma'}$ represents the most general four fermion covariant interaction in curved space  \cite{PhysRevB.59.7140}.
{\ba\label{eq:interaction.tensor.space}
T^{\sigma_1\sigma_2}_{\sigma_3\sigma_4}
&=&  g_1\;\delta^{\sigma_1}{}_{\sigma_3}\;\delta^{\sigma_2}{}_{\sigma_4}
+ g_2\;(\Gamma^5)^{\sigma_1}{}_{\sigma_3}\;(\Gamma^5)^{\sigma_2}{}_{\sigma_4}
+g_3\;(\Gamma^a)^{\sigma_1}{}_{\sigma_3}\;(\Gamma_a)^{\sigma_2}{}_{\sigma_4}\cr
&+&
g_4\;  
(\Gamma^a\,\Gamma^5)^{\sigma_1}{}_{\sigma_3}\;
(\Gamma_a\,\Gamma^5)^{\sigma_2}{}_{\sigma_4}
-\frac{g_5}{4}\; 
\left([\Gamma^a,\Gamma^b]\,\Gamma^5\right)^{\sigma_1}{}_{\sigma_3}\;
\left([\Gamma_a,\Gamma_b]\,\Gamma^5\right)^{\sigma_2}{}_{\sigma_4}\,.
\ea
We see that it} is written completely in terms of Dirac matrices, and depends linearly on a set of five coupling constants $g_1,\dots,g_5$, which we assume are small.

We can now obtain the Hamiltonian resulting from  \eqref{eq:dirac.action} and  \eqref{eq:interaction.action}, it takes the form 
\small
\bea\label{eq:hamiltonian}
H &=&
\sum_{\alpha m}\!
\int \!d^2\!k\;\omega_{m}(k)\,
c_{\alpha m \vec k}^\dagger
c_{\alpha m \vec k}\\
&+&
\sum_{ \underset{m_1m_2m_3m_4}{\mbox{\tiny$\alpha_1 \alpha_2\alpha_3\alpha_4$\normalsize}}}
\!\!\!
\int d^2\!k\,d^2\!k' d^2\! q\;\,
t^{\alpha_1 m_1({\vec k}\!+\!\vec q);\alpha_2 m_2({\vec k}'\!-\!\vec q)}_
{\alpha_3 m_3{\vec k};\alpha_4 m_4{\vec k}'}\;  
c_{\alpha_1 m_1 (\vec k+\vec q)}^\dagger\;c_{\alpha_2 m_2 (\vec k'-\vec q)}^\dagger
c_{\alpha_3 m_3 {\vec k}}\;c_{\alpha_4 m_4 {\vec k}'}\,.
\nonumber
\eea
\normalsize
Here  the frequencies $\omega_m(k)$ play the role of dispersion relations for each fermionic mode. 
On the other hand the new tensor $t^{\alpha_1 m_1{\vec k}_1;\alpha_2 m_2{\vec k}_2}_
{\alpha_3 m_3{\vec k}_3;\alpha_4 m_4{\vec k}_4}$ contains the information about the interaction strengths among different modes, and it results from contracting the position space interaction tensor $T^{\bar\sigma\bar\sigma'}_{\sigma\sigma'}$ with  integrals in the $z$ direction of quartic products of the free fermionic eigenstates $\psi_{\alpha m\vec k }(z)$. 

\bigskip

The explicit forms of  the interaction tensors $T^{\bar\sigma\bar\sigma'}_{\sigma\sigma'}$  and $t^{\alpha_1 m_1{\vec k}_1;\alpha_2 m_2{\vec k}_2}_{\alpha_3 m_3{\vec k}_3;\alpha_4 m_4{\vec k}_4}$. as well as that of the aforementioned quartic integrals, is not relevant for the moment. For the details of the derivation of \eqref{eq:hamiltonian}  we refer the reader to Appendix \ref{app:hamiltonian.theory}.

\bigskip

With equation \eqref{eq:hamiltonian}, we have succeded in re-writing the bulk dynamics as that of a second quantized Hamiltonian for a two dimensional multi-fermion system in which the index $m$ denotes fermion species. Since the coupling in the bulk is assumed to be weak, we can safely rely on the Landau description of the Fermi liquid.

Assuming that the ground state of the Hamiltonian \eqref{eq:hamiltonian} is characterized by a certain set of occupation numbers $N_{\alpha m \vec k}$, then the excitations can be described by their variations  $\delta N_{\alpha m \vec k}$. The grand canonical energy of an excitation is then written as the Landau formula
\small
\be \label{eq:grand.canonica.perturbations}
\delta \Omega(T,\mu)\!=\!
\int d^2\vec k\,\sum_{\alpha m  } \,\epsilon_{m}( k)\delta N_{\alpha m\vec k }
+ 
\frac12
\int \!d^2 \vec k\,  d^2 \vec k'\!\!\sum_{\alpha m \alpha' m'}
\,f_{\alpha m \alpha' m'}(\vec k,\vec k')\, 
\delta N_{\alpha m\vec k}\, \delta N_{\alpha' m'\vec k'}\,,
\ee 
\normalsize
where $f_{\alpha  m\bar \alpha  \bar m}(\vec k,\vec {\bar k})$ is the so-called ``interaction function'' which is obtained from the tensor $t^{\alpha_1 m_1{\vec k}_1;\alpha_2 m_2{\vec k}_2}_
{\alpha_3 m_3{\vec k}_3;\alpha_4 m_4{\vec k}_4}$, and the quasiparticle dispersion relation $\epsilon_m(k)$ takes the form 
\be
\label{eq:dispersion.relation}
\epsilon_m(k)=\omega_m(k)+\sum_{ \alpha' m'}\int \!d^2 \vec k'\,   f_{\alpha m \alpha' m'}(\vec k,\vec k')\, N_{\alpha' m'\vec k'}\,.
\ee
For future use, notice that the quasiparticle dispersion relation equals the frequency plus corrections of first order in the coupling constants.

For the most general covariant quartic perturbation \eqref{eq:interaction.action}, the interaction function takes a particularly simple angular dependence in the momentum plane. Indeed, if we write $\vec k=k\,(\cos\theta,\sin\theta)$ then it can be decomposed in only three Fourier modes, as
\be\label{eq:Fourier.decomposition}
 f_{\alpha m;\alpha' m'}(\vec k, {\vec k}') = 
 f_{\alpha m k;\alpha' m' k'}^0 + 
\cos(\theta-\theta')\;f_{\alpha m k;\alpha' m' k'}^c + 
\sin(\theta-\theta')\;f_{\alpha m k;\alpha' m'{ k}'}^s\,,
\ee
where the constant, sine and cosine coefficients depend on a reduced subset of the components of the tensor  $t^{\alpha_1 m_1{\vec k}_1;\alpha_2 m_2{\vec k}_2}_
{\alpha_3 m_3{\vec k}_3;\alpha_4 m_4{\vec k}_4}$, or in other words on the position space interaction tensor $T^{\bar\sigma\bar\sigma'}_{\sigma\sigma'}$ contracted with integrals of the fermionic states $\psi_{\alpha m\vec k }(z)$.

\bigskip

The details of the construction of the Landau description starting with the underlying Hamiltonian \eqref{eq:hamiltonian}, as well as the explicit form of the interaction function \eqref{eq:Fourier.decomposition} in terms of the interaction tensor and wavefuntion integrals, are presented in Appendix \ref{app:Landau.description}.

\bigskip

We can now make use of the Pomeranchuk technique for the above defined two dimensional multicomponent Landau Fermi liquid, in order to diagnose instabilities arising from an anisotropic deformation of its Fermi surface.

We start by decomposing the deformation on the occupation numbers, that charactherize the excitations of the Fermi liquid, in the form
\small
\be
\delta N_{\alpha m\vec k}={\sf H}\big{(}\!-\!\epsilon_{ m}(k) + \delta g_{\alpha m}(\vec k)\big{)} - {\sf H}\left(\!- \epsilon_{ m}(k) \right)
=
\delta\big{(}\!-\!\epsilon_m(k) \big{)}\; \delta g_{\alpha m}(\vec k) 
+\frac12\;\delta'\big{(}\!-\epsilon_m(k) \big{)}\;\delta g_{\alpha m}(\vec k)^2  +\dots 
\label{eq:occupation.numbers.perturbation}
\ee
\normalsize
where ${\sf H}(\cdot)$ is the Heavyside unit-step function, and $\delta g_{\alpha m}(\vec k)$ are arbitrary functions of the momentum characterizing the excitation. When plugged back into \eqref{eq:grand.canonica.perturbations} the delta functions force the evaluation of the expression on the Fermi momentum $k_F^{m}$ defined by $\epsilon_m(k_F^{m}) =0$. The remaining angular integrals result on a grand canonical energy that is a quadratic form in the Fourier components 
$\delta g_{\alpha m}{}_n^{(c,s)}$ of the parameters $\delta g_{\alpha m}(\vec k)|_{k=k_F^m}= \sum_{n=0}^\infty \left(\delta g_{\alpha m}{}_n^c\;\cos(n\theta) +\delta g_{\alpha m}{}_n^s\;\sin(n\theta)\right)$.
{Notice that, since the interaction function \eqref{eq:Fourier.decomposition} has only three Fourier components, the only modes that depend on the couplings (and hence could lead to negative minors in the quadratic form) are $\delta g_{\alpha m}{}_0^{c}$ and $\delta g_{\alpha m}{}_1^{(c,s)}$. The first one is isotropic, while the remaining two completely break the rotational symmetry.}

If the grand canonical energy of an excitation is negative, the system decreases its energy by creating more such excitations. To avoid such instability, we need to impose that the aforementioned quadratic form is positive definite. This is guaranteed provided all the minors of the quadratic kernel are positive, or in other words 
\bea
\left|\frac{k_F^{ m}}{v_F^{ m}}
\left(\delta_{\alpha m \alpha' m'} +
2\,\pi\,\frac{k_F^{ m'}}{v_F^{m'}}\;
f^0_{\alpha m\alpha'm' }\right)
\right|_{M\times M}&>&0\quad,\quad\forall M\in \mathbb{N}
\label{eq:Pomeranchuk.conditions1} 
\\
\label{eq:Pomeranchuk.conditions2} 
\left|\frac{k_F^{ m}}{v_F^{ m}}
\left(\delta_{\alpha m \alpha' m' }+\pi
\frac{k_F^{ m'}}{v_F^{ m'}}\,\left({f^c_{\alpha m \alpha' m' }}+i{f^s_{\alpha m \alpha' m' }}\right) \right)\,
\right|_{M\times M}
&>&0\quad,\quad\forall M\in \mathbb{N}
\eea
where $v_F^{ m}=d\epsilon_m(k)/dk|_{k=k_F^m}$ are the Fermi velocities and $f^{(0,c,s)}_{\alpha m \alpha' m' }$ are the Landau parameters, given by the Fourier components  in \eqref{eq:Fourier.decomposition} evaluated at the corresponding Fermi momenta. 
Notice that, provided the Fermi velocities are positive, the quotient ${k_F^{ m}}/{v_F^{ m}}$ can be removed from the prefactors. Moreover, we can perturbatively replace ${k_F^{ m}}/{v_F^{ m}}$ in the parenthesis by its free version, defined by $\omega_m(k_f^{m \,{\sf free}})=0$ and $v_F^{ m\,{\sf free}}=d\omega_m(k)/dk|_{k=k_F^{m\,{\sf free}}}$.

{When some of the minors in \eqref{eq:Pomeranchuk.conditions1} is negative, a linear combination of the modes $\delta g_{\alpha m}{}_0^{c}$ causes a net decrease in the energy, resulting in an instability. As mentioned above, such modes are isotropic and therefore the corresponding instability would be a Stoner one, {\em i.e.} an instability with respect to adding or removing particles from the ground state. On the other hand, if the negative minor shows up in \eqref{eq:Pomeranchuk.conditions2}, the unstable direction correspond to a linear combination of the modes  $\delta g_{\alpha m}{}_1^{(s,c)}$, and the resulting deformation completely breaks rotational invariance. It is tempting to imply that such symmetry breaking pattern is inherited by the phase that stabilizes the system after the phase transition.}

A review of Pomeranchuk method, including the details of the calculations resulting in formulas \eqref{eq:Pomeranchuk.conditions1}-\eqref{eq:Pomeranchuk.conditions2}, can be seen in Appendix \ref{app:Pomeranchuk.method}.

\section{Summary of the method}
\label{sec:application.example}
In summary, in order to test the anisotropic Pomeranchuk instabilities of an holographic Fermi liquid, we need to 
\begin{enumerate}

{ 

\item Obtain the fermionic modes $\psi_{\alpha m k}(z)$ in the expansion 
\eqref{eq:fermion.decomposition}, that we read from \eqref{eq:fermion.decomposition.final} and  \eqref{eq:fermion.decomposition.annihiltion.operator}), and their frequencies $\omega_m(k)$. 
This is done by solving a Schr\"odinger-like equation with potential \eqref{eq:potential} in the holographic direction for a function $\phi_{mk}(z)$. Notice that we only need the modes with very small frequency, since we will evaluate them at the free Fermi momentum $k_F^{m\,{\sf free}}$ where the frequency vanishes.
\label{item:fermionic.modes}
}

\item Calculate the free Fermi velocities as $v_F^{m\,{\sf free}}=d\omega_F^{m}/dk|_{k=k_F^{m\,{\sf free}}}$. Incidentally, this is why we need modes with non-vanishing small frequency, even if just those with vanishing frequency contribute to the quantities evaluated at the free Fermi momenta.\label{item:fermi.velocities}

{

\item Integrate a particular set of quartic products of  fermionic modes in the holographic direction given in \eqref{eq:fermion.integrals.all}, in order to go from the coupling tensor in spin indices 
 $T^{\bar\sigma\bar\sigma'}_{\sigma\sigma'}$ given by 
to $t^{\alpha_1 m_1{\vec k}_1;\alpha_2 m_2{\vec k}_2}_{\alpha_3 m_3{\vec k}_3;\alpha_4 m_4{\vec k}_4}$ 
and then to the Landau parameters $f^{( c,s)}_{\alpha m \alpha' m' }$.\label{item:Integration.of.modes}
}
\item Check whether all the minors in \eqref{eq:Pomeranchuk.conditions1}-\eqref{eq:Pomeranchuk.conditions2} are positive. Notice that each minor of $M\times M$ has degree $M$ on the interaction function, which is linear in the coupling constants. This implies that each condition imposes a polynomial restriction of order $M$ in coupling space.
\label{item:Pomeranchuk.condition}
\end{enumerate}
At any point in coupling space at which condition \ref{item:Pomeranchuk.condition} is not satisfied, the quadratic form has a negative eigenmode, implying that the grand canonical energy  is decreased by the corresponding excitation. This results in an instability of the system. {If the negative minor appears in \eqref{eq:Pomeranchuk.conditions1}, the resulting instability is isotropic, while if it shows up in \eqref{eq:Pomeranchuk.conditions2}, it breaks completely the rotational symmetry.}

\newpage

\section{Application to the electron star}

We applied the above steps to the specific example of the zero temperature electron star background of \cite{Hartnoll:2010gu}. The electron star represents the holographic dual to the ground state of a highly degenerate system of fermions at zero temperature. It is constructed by coupling the gravitational and electromagnetic degrees of freedom in the bulk to a perfect fluid representing the fermions, whose equation of state is obtained in the Thomas-Fermi approximation. A summary of the electron star solution is given in Appendix \ref{app:electron.star}. 

In order to obtain the fermionic modes and calculate the Fermi momenta and velocities as required by points \ref{item:fermionic.modes} and \ref{item:fermi.velocities}, we used a Wentzel-Kramers-Brillouin (WKB) approximation on the spinorial perturbation. This is consistent with the use of the Thomas-Fermi approximation for the bulk fluid. The relevant details of the WKB method are presented in Appendix \ref{app:WKB}.

\begin{center}
\begin{figure}[H]
\begin{center}
\includegraphics[width=6cm]{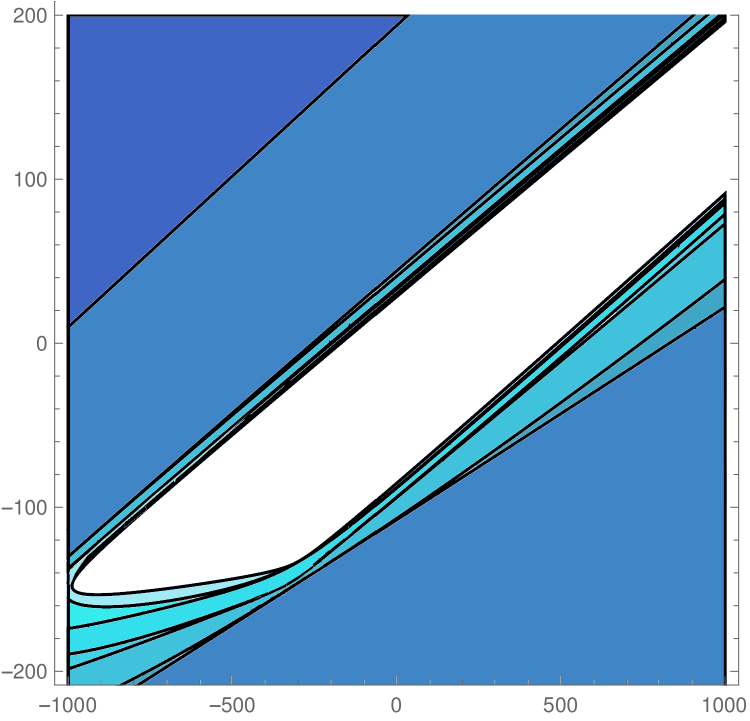}\quad\quad
\includegraphics[width=6cm]{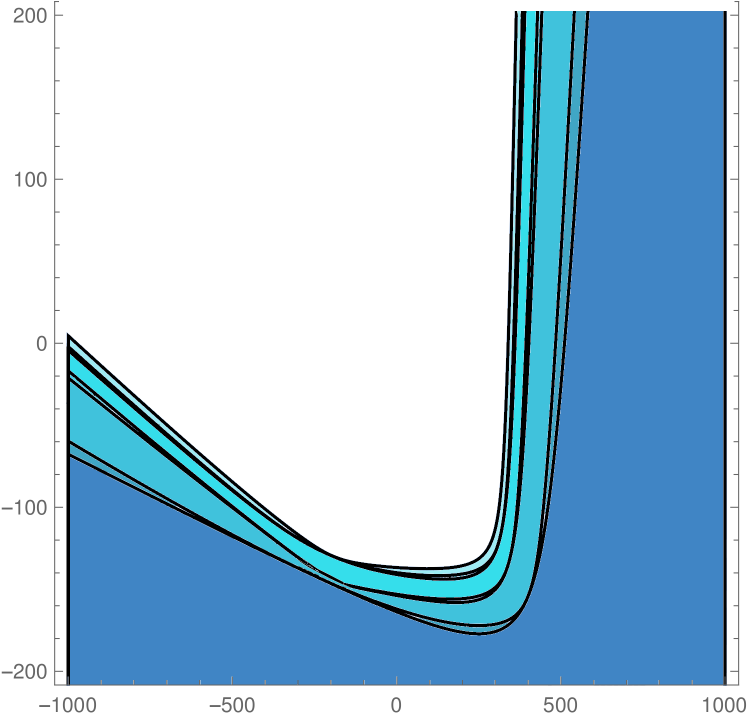}
\\~\\
~
\!\!\!\!\!\!\!\!\!\!\!\!\!\!\!\!\!\!\!\!\!\!\!
\!\!\!\!\!\!\!\!\!\!\!\!\!\!\!\!\!\!\!\!\!\!\!
\!\!\!\!\!\!\!\!\!\!\!\!\!\!\!\!\!\!\!\!\!\!\!
\includegraphics[width=6.2cm]{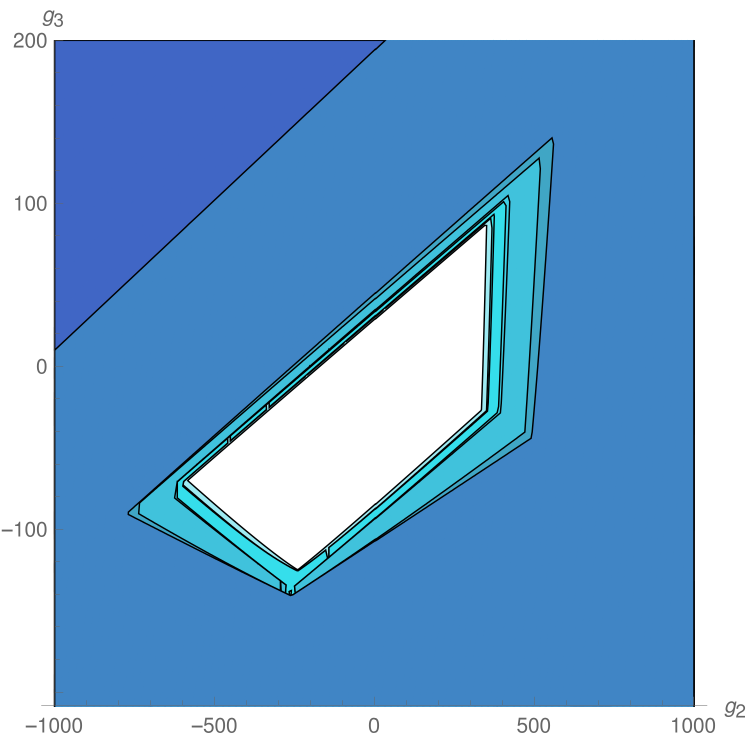} \quad\quad \includegraphics[width=1.7cm]{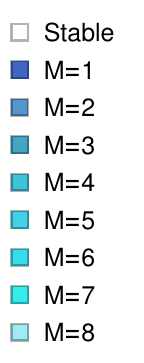}
\end{center}
\caption{Construction of the unstable regions in the coupling plane $g_2$-$g_3$. The plots on the top correspond to the unstable regions in equation \eqref{eq:Pomeranchuk.conditions1} (left) and \eqref{eq:Pomeranchuk.conditions2} (right). The plot on the bottom combines both equations. The colors denote the order $M$ of the first negative minor.
}
\label{fig:construction}
\end{figure}
\end{center}

\begin{center}
\begin{figure}[H]
\includegraphics[width=4.7cm,height=4.75cm]{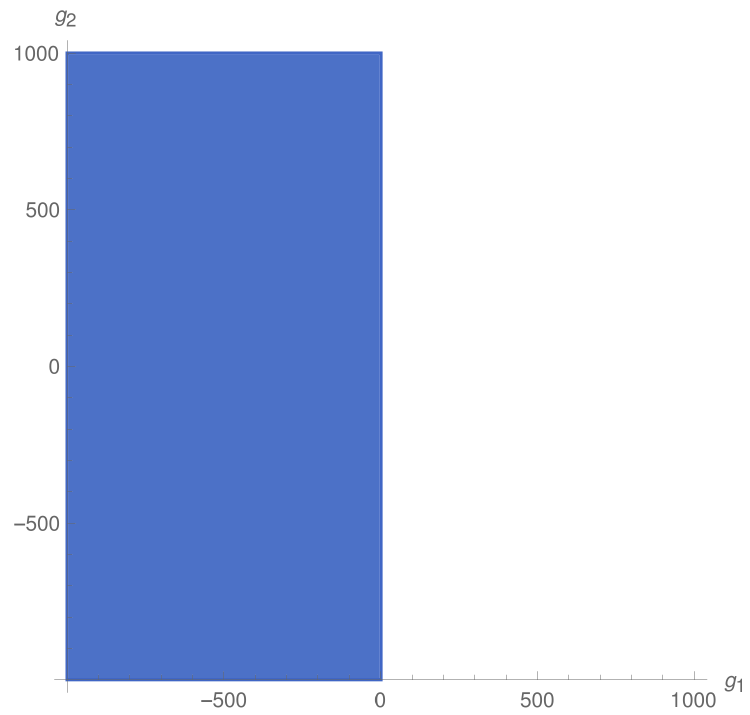}\quad\!\!\!\!
\includegraphics[width=4.7cm]{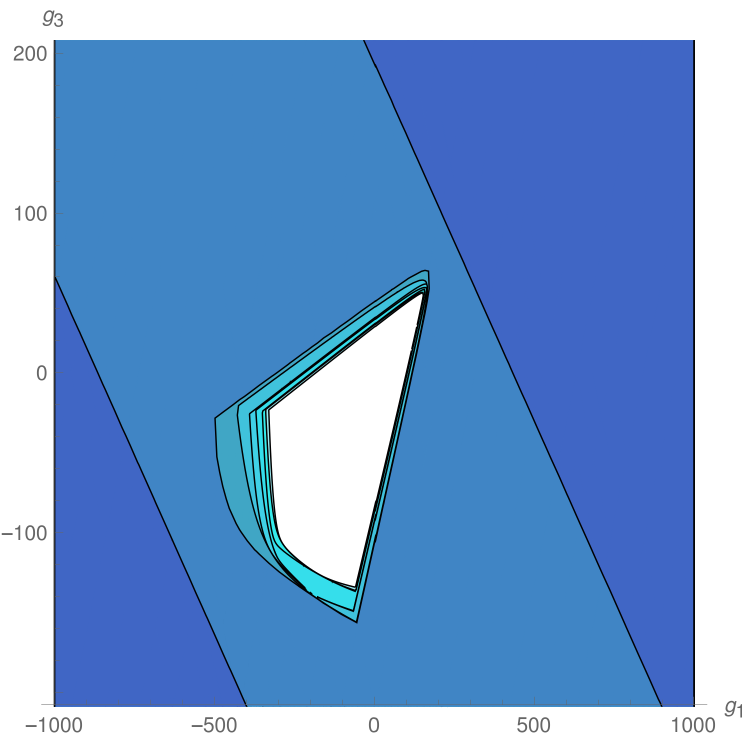}\quad
\includegraphics[width=4.75cm]{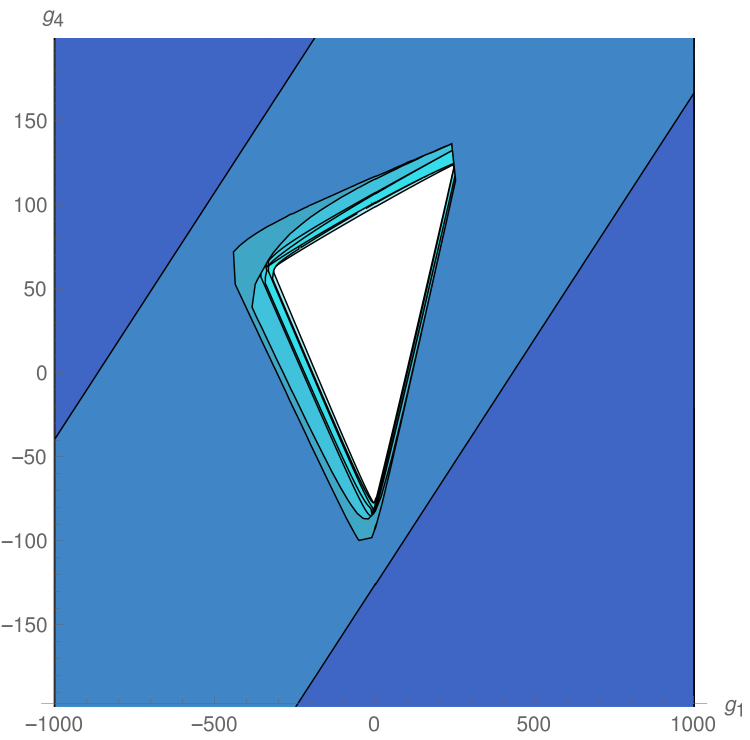} 
\\~\\
\includegraphics[width=4.7cm]{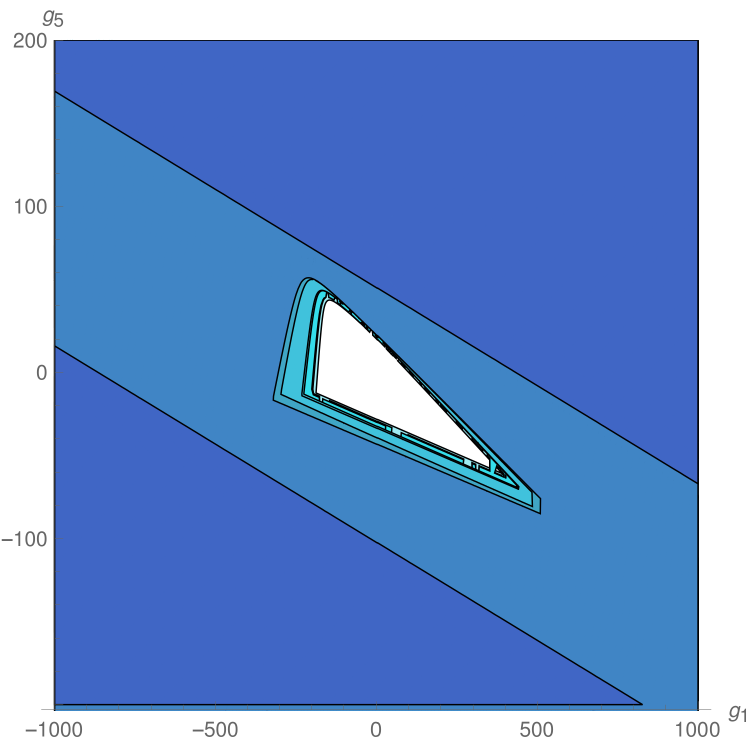}\quad\!\!\!\!
\includegraphics[width=4.7cm,height=4.75cm]{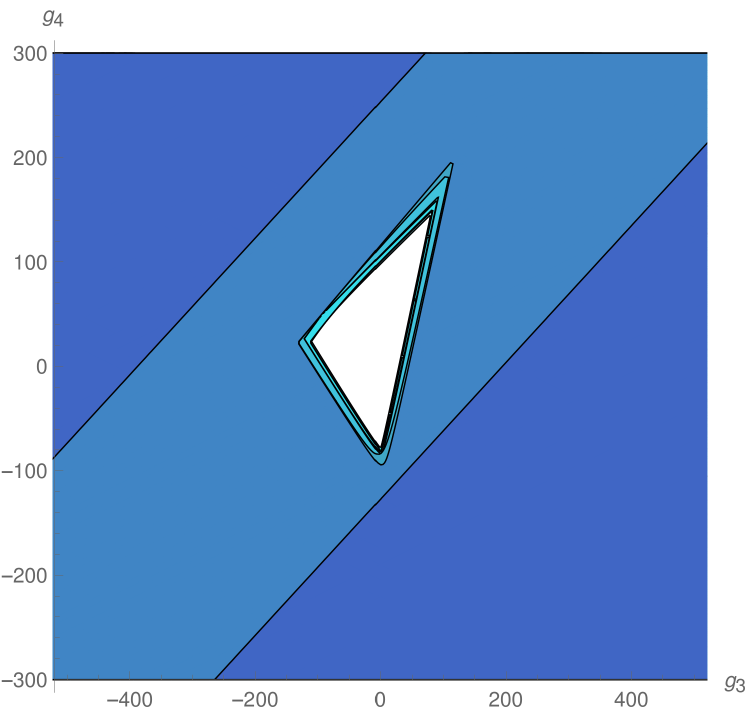}\quad
\includegraphics[width=4.7cm,height=4.72cm]{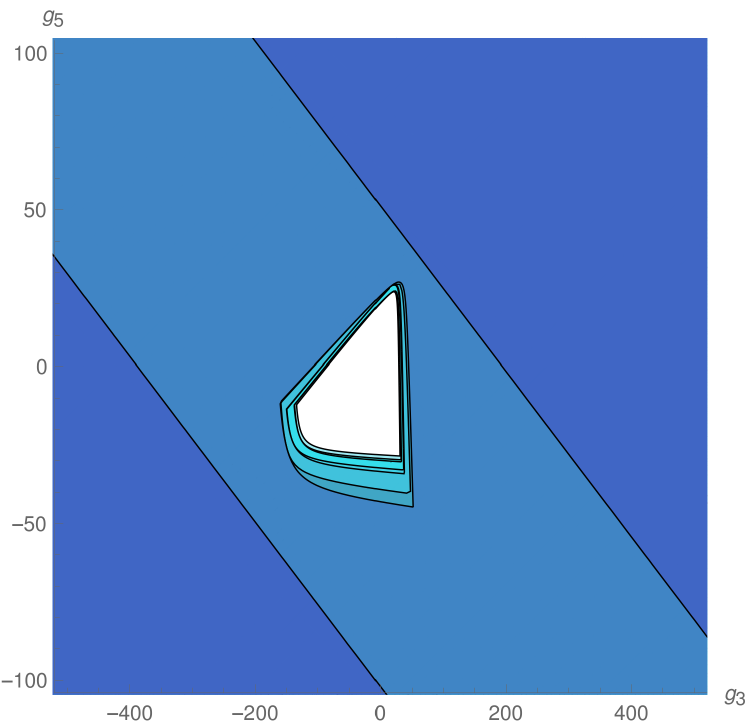} 
\\~\\
\includegraphics[width=4.75cm,height=4.75cm]{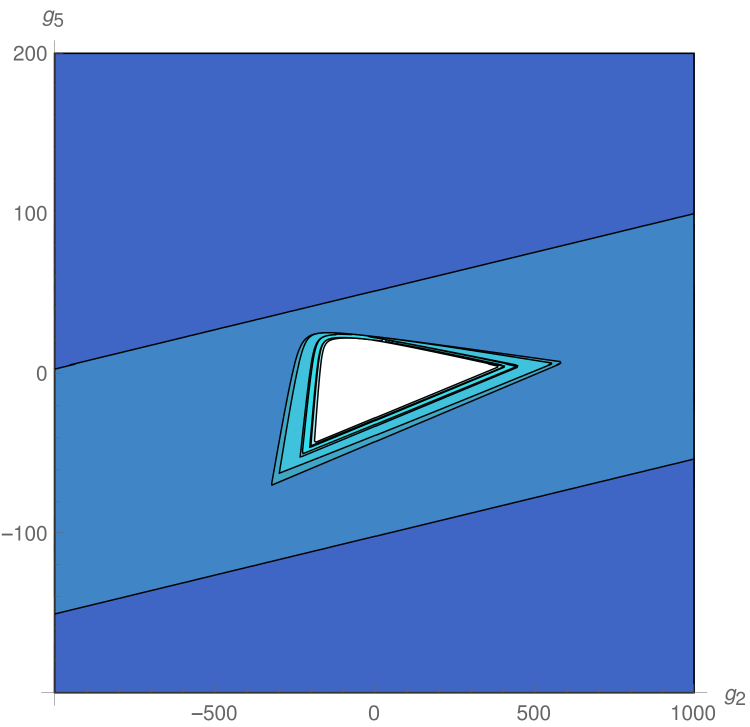}\quad\!\!\!\!\!
\includegraphics[width=4.77cm]{gm3vsgm2.pdf}\quad\!\!\!
\includegraphics[width=4.75cm,height=4.77cm]{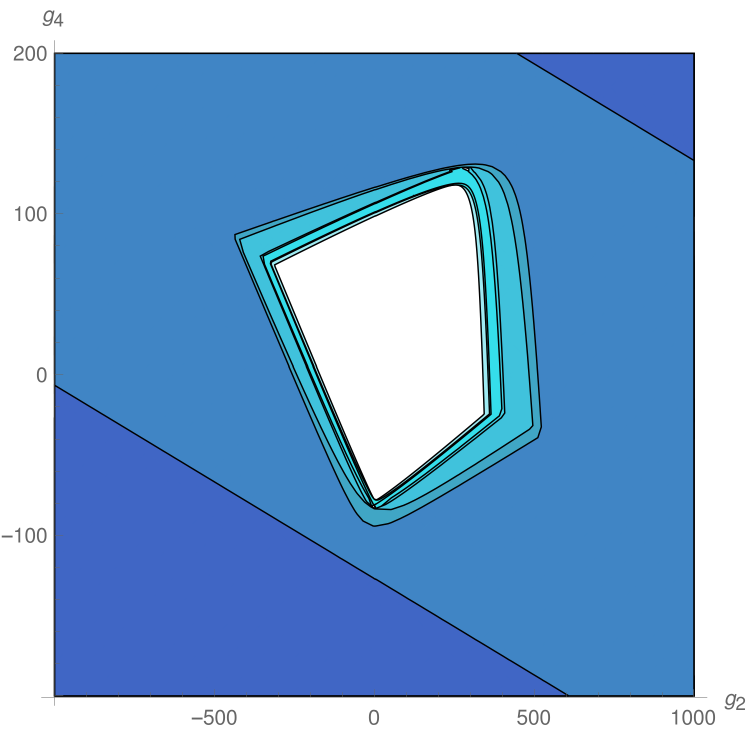}
\\~\\
\includegraphics[width=4.75cm,height=4.75cm]{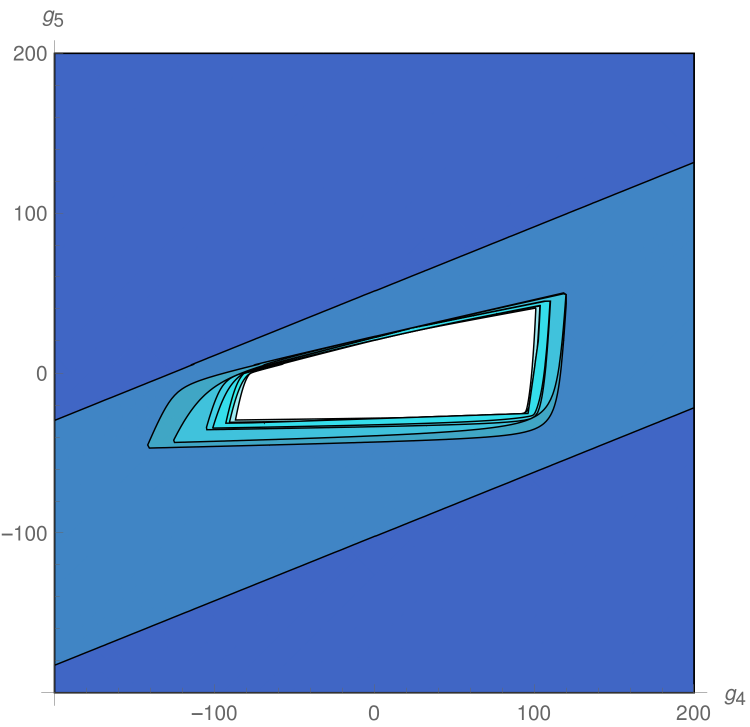}\quad\ 
\includegraphics[width=1.7cm]{Legend.pdf}
\caption{Phase diagrams showing the unstable regions in the five-dimensional coupling space  $g_{1}$, $g_{2},g_{3},g_{4},g_{5}$. Each plot corresponds to one of the ten two-dimensional coupling planes. We see that there is an island of stability around the origin. The lighter regions become unstable at higher $M$. The plots correspond to $\hat{m}=0.4$, $\lambda=2$ and $\gamma=100$ (see Appendix \ref{app:electron.star} for details on these parameters).
}
\label{fig:phase.diagrams}
\end{figure}
\end{center}

{We explored the phase diagram by turning on the coupling  constants $g_1,\dots,g_5$ by pairs. Sitting at different values of the non-vanishing couplings on every coupling plane, we evaluated the minors in equations \eqref{eq:Pomeranchuk.conditions1}  and \eqref{eq:Pomeranchuk.conditions2} order by order. For example, on the plane $g_2$-$g_3$, we obtained the plots at the top of Fig.\ref{fig:construction}. The regions are painted with a color that depends on the order $M$ of the first unstable minor. The plot on the top left correspond to instabilities that preserve rotational invariance, while that on the top right denotes instabilities that break it completely. These plots were combined on the one at the bottom.} The resulting unstable regions of the phase diagram can be seen in Fig. \ref{fig:phase.diagrams}, in which all different coupling planes are presented. We see that as $M$ grows (and so does the degree of the polynomial in point \ref{item:Pomeranchuk.condition} above) the region in coupling space where the system is stable gets smaller, converging to an isolated wide island around the origin, which is simply connected.

\bigskip

{In order to interpret our results from the boundary perspective, we recall that the electron star describes a holographic strongly coupled theory whose ground state has a fermionic condensate \cite{Hartnoll:2010gu}. Classical spinor perturbations on the bulk then correspond to fermionic excitations on the boundary theory \cite{Hartnoll:2011dm}. 

The interesting point is that, as we decompose the bulk spinor into $z$-modes, the set of Fermi momenta we obtained for them corresponds exactly to those defining the ``holographically smeared Fermi surface'' in reference \cite{Hartnoll:2011dm}. It is natural then to interpret collective bulk modes, made by deforming the Fermi surface of each bulk $z$ mode, as corresponding to collective excited states on the boundary theory.  These ``deformations of the smeared Fermi surface'' become unstable when a bulk Pomeranchuk instability shows up, leading to an instability of the boundary state that could easily be interpreted as a boundary Pomeranchuk instability. }

\section{Conclusions and outlook}
\label{sec:conclusions}
We developed a general method to study anisotropic fermionic instabilities on a strongly coupled Fermi liquid via the AdS/CFT duality. It entails to perform a Pomeranchuk analysis on the dual bulk fermion, adapting the formalism to a curved space-time with a planar slice. The key step is to rewrite the single fermion in the bulk in terms of its modes in the holographic direction, resulting on a multi-component systems of lower dimensional fermions which move in the planar slice. By analyzing the minors of a quadratic form, we are able to detect when the fermionic modes become unstable under anisotropic deformations of their Fermi surface. The resulting phase transition breaks the rotational symmetry in the planar slice, and consequently in the boundary theory, potentially leading to a nematic phase.

We kept the formalism as general as possible, being suitable to be applied to {many Euclidean-invariant boundary theories. This includes the zero temperature electron star \cite{Hartnoll:2010gu,Hartnoll:2011dm}, the finite temperature electron star \cite{Puletti:2010de,Hartnoll:2010ik}, as well as other examples\footnote{Since the method relays on the existence of a bulk Fermi surface, to adapt it to backgrounds such like the holographic metal \cite{Lee:2008xf, Liu:2009dm,Faulkner:2010da,Faulkner:2009wj}, the holographic superconductor \cite{Hartnoll_2008}, the magnetically charged black-hole \cite{Albash_2008}, or the Lifshitz geometry \cite{Kachru:2008yh}, one would need to work in the limit in which a bulk fermion condensate would not backreact. This limit can be subtle, and it is hard to reconcile with the validity conditions for the WKB method, so the fermion wave functions should be solved using a different approximation\label{foot}.}}. We applied the method to the zero-temperature electron star background, being able to identify the unstable region on a five dimensional coupling space.
 
The method can be extended to more general situations relevant to condensed matter systems. Some examples of possible future directions are:
\begin{itemize}
\item Finite doping: the inclusion of a doping axis should be straitforward, following the lines described in \cite{Kiritsis2016}.
\item Finite magnetic field: the inclusion of a magnetic field would require a magnetically charged background \cite{Albash_2008}\footnote{See footnote \ref{foot}.\label{foot2}}, and it would modify the bulk fermion coupling. Also, the Pomeranchuk analysis needs to be adapted \cite{dos}.
\item Finite temperature fluid: as it stands, the method can be applied to the background of 
\cite{Puletti:2010de, Hartnoll:2010ik} in which the temperature is included via the presence of a horizon, while the bulk fluid is approximated as a finite temperature Fermi liquid. Including the effect of temperature in the bulk matter would modify the background, as well as the Pomeranchuk analysis \cite{uno}.

\item Lifshitz scaling: considering Lifshitz geometries \cite{Kachru:2008yh}  would require a rewriting of the asymptotic conditions on the near boundary fermion, but the rest of the analysis remains mostly unchanged$^{\ref{foot2}}$.
\item Lattice fermions: describing an underlying lattice in the holographic setup would imply the inclusion of a lattice-like metric \cite{Horowitz2012} or momentum relaxation \cite{Andrade2014}, and a modification of the Pomeranchuk analysis according to the lines of \cite{PhysRevB.78.115104}.
\end{itemize} 
These are only some of the situations to which the method can be extended, we think that they deserve further investigation. 

{As a  last remark, one could wonder whether Pomeranchuk instabilities are evident in physical quantities as we approach the edge of the unstable region\footnote{We would like to thank the referee for rising this point.}. In a weakly coupled Fermi liquid, standard Landau theory arguments show that Landau parameters appear into some response functions. This happens for example for the compressibility or the spin susceptibility. The expressions are such that these functions become infinite when the instability is triggered by the correct Landau parameter. 

Recalling that, according to Kubo formulae, response functions are written in terms of two-point correlators of local quantities, we can calculate them in holography. This is done by perturbing the background fields with the right IR boundary conditions and then evaluating the quotient of the leading and subleading parts of the perturbations at the UV. However, it is evident that in such calculation the fermionic couplings $g_1,\dots,g_5$ would only enter at higher loops. Since there is no reason to assume that bulk perturbation theory would break down close to the edge of the unstable region, the natural conclusion is that, in the strongly coupled boundary theory, the response functions are not sensitive to the instability.}

\section*{Aknowledgements}
The authors thank Pablo Rodr\'\i guez Ponte and Ignacio Salazar Landea for relevant contributions during the early stages of this work. We also thank Carlos Lamas for helpful comments. This work has been funded by the CONICET grants PIP-2017-1109, PIP-2015-0688 and PUE 084 ``B\'usqueda de Nueva F\'\i sica'', and UNLP grants PID-X791, PID-11/X910. 
\newpage
\appendix

~

\section{Fermionic states in a holographic background}
\label{app:free.fermionic.states}
In this Appendix, we work out a basis of free fermionic modes in a charged planar holographic background. We do so by writing the background in terms of vielbein and spin connection (section \ref{app:free.fermionic.states.background}), deriving the form of the corresponding Dirac equation (section \ref{app:free.fermionic.states.dirac}), separating it into spin components and Fourier modes in the planar and time directions  (section \ref{app:free.fermionic.states.dirac.separation}), and obtaining an effective Schr\"odinger equation that describes the fermion profile in the holographic direction (section \ref{app:free.fermionic.states.effective.schroedinger}). We write the solution (section \ref{app:free.fermionic.states.general.form}) and notice that boundary conditions would generically imply a quantization of the frequencies as functions of the momenta (section \ref{app:free.fermionic.states.general.form.quantization}), resulting in a bunch of $2+1$ dimensional dispersion relations. We finally obtain the normalization constants, and check that with the so defined basis the resulting second quantized theory satisfy correctly normalized commutation relations (section \ref{app:free.fermionic.states.general.form.quantization}).
\subsection{The background}
\label{app:free.fermionic.states.background}
We work with a generic planar asymptotically AdS charged background, with the form
\bea
\label{eq:background.detailed}
G &=& L^2\left(
-f \, dt^2+g \, dz^2 +\frac{d\vec x^2}{z^2}
\right)=\eta_{AB}\,\omega^A\,\omega^B \,,\cr
A&=&
h  \,dt=A_{A}\;\omega^{A}\,.
\eea
Here $f$, $g$ and $h$, and consequently $\omega^A$ and $A^A$, are functions of $z$.
Since we want to couple fermions to the present background, we need the explicit form of the vielbein and dual vector basis $\{\omega^A, e_A\}$ that we wrote in the second equality. In terms of the planar  indices \scriptsize $A$ \normalsize $=\ut, \ux, \uy, \uz$, they read
\be\label{eq:vielbein}
\begin{array}{lcl}
\omega^{\ut}\equiv L\,\sqrt{f}\;dt\,\qquad&,&\qquad e_{\ut}\equiv \frac{1}{L\,\sqrt{f}}\;\partial_t\,,\cr
\omega^{\ux}\equiv \frac{L}{z\,}\;dx\,\qquad&,&\qquad e_{\ux}\equiv \frac{z\,}{L}\;\partial_x\,,\cr
\omega^{\uy}\equiv \frac{L\,}{z}\;dy\,\qquad&,&\qquad e_{\uy}\equiv \frac{z}{L\,}\;\partial_y\,,\cr
\omega^{\uz}\equiv L\,\sqrt{g}\;dz\,\qquad&,&\qquad e_{\uz}\equiv \frac{1}{L\,\sqrt{g}}\;\partial_z\,.
\end{array}
\ee
This allows us to calculate the non-zero components of the spin connection $\{\omega^A{}_B\}$ by imposing metricity $\omega_{AB}=-\omega_{BA}$, and torsionless $\;d\omega^A + \omega^A{}_B\wedge\omega^B =0\;$ conditions. They take the form
\bea
\omega^{\ut}{}_{\uz}&=&-\omega_{{\ut}{\uz}}=+\omega_{{\uz}{\ut}}= 
\frac{(\ln f)'}{2\,\sqrt{g}}\;\omega^{\ut}\,,\cr
\omega^{\ux}{}_{\uz}&=&+\omega_{{\ux}{\uz}}=-\omega_{{\uz}{\ux}}= -\frac{1}{z\sqrt{g}}\;\omega^{\ux}\,,\cr
\omega^{\uy}{}_{\uz}&=&+\omega_{{\uy}{\uz}}=-\omega_{{\uz}{\uy}}=
-\frac{1}{z\sqrt{g}}\;\omega^{\uy}\,.
\label{eq:spin.connection}
\eea
This allows us to write the Dirac equation for a charged fermion in the present background, as we do in the next subsection.
\newpage
\subsection{The Dirac equation}
\label{app:free.fermionic.states.dirac}
\subsubsection{Separation of variables and spin components}  
\label{app:free.fermionic.states.dirac.separation}
We consider a four-component Dirac spinor $\Psi$ with charge $q$ under the $U(1)$ gauge field coupled to gravity through the covariant derivative
\be\label{eq:covariant.derivative}
\slashed {\cal D}\,\Psi \equiv \Gamma^A\,{\cal D}_A\,\Psi =
\Gamma^A\left( e_A(\Psi) + \frac{i}{2}\,\omega^{BC}{}_A\,\Sigma_{BC}\,\Psi - i\, q\, A_A\, \Psi\right)\,,
\ee
where the Dirac gamma-matrices obey $\{\Gamma^A,\Gamma^B\} = 2\,\eta^{AB}$, and $\Sigma_{AB}\equiv [\Gamma_A, \Gamma_B]/{4\,i}$
are the generators in the spinorial representation of the local Lorentz group in $3+1$ dimensions.
We use along the paper the following representation for the gamma matrices
\be\label{eq:gamma.matrices}
\Gamma^{\ut}\equiv\left(\begin{array}{cc}i\,\sigma_1&0\\0&i\,\sigma_1\end{array}\right)\,,
\quad\quad
\Gamma^{\ux}\equiv\left(\begin{array}{cc}-\sigma_2&0\\0&\sigma_2\end{array}\right)\,,
\quad\quad
\Gamma^{\uy}\equiv\left(\begin{array}{cc}0&\sigma_2\\ \sigma_2&0\end{array}\right)\,,
\quad\quad
\Gamma^{\uz}\equiv\left(\begin{array}{cc}\sigma_3& 0\\ 0&\sigma_3\end{array}\right)\,,
\ee
where $\{\sigma_1,\sigma_2, \sigma_3\}$ are the Pauli matrices. 

The free fermionic modes satisfy the Dirac equation 
\be\label{eq:dirac.equation}
(\slashed {\cal D} - m)\,\Psi =0\,.
\ee
To solve it, we find convenient to work in momentum space
\be\label{eq:fermion.decomposition.first}
\Psi_{\omega\vec k}(t,\vec x,z)=\Psi_{\omega\vec k}(\vec x,z)e^{-i\omega t}= \frac1{{\cal N}_{\omega\vec k}}\; \frac{z}{f(z)^\frac{1}{4}}\,e^{i(\vec k\cdot\vec  x-\omega t)}\,\psi_{\omega \vec k}(z)\,,
\ee 
where $\omega$, $\vec k$ are the energy and momentum along the $xy$ plane, ${\cal N}_{\omega\vec k}$ is a normalization constant, and the factor $z/f(z)^{1/4}$ has been included for later convenience. We can use rotational invariance to refer the momentum to the $x$-axis, as
\ba\label{eq:rotation.matrices.vector}
R[\theta]\left(
\begin{array}{c}
k_x\\
k_y
\end{array}\right)=
\left(
\begin{array}{c}
k\\
0
\end{array}\right)\,,
\qquad{\rm with}\qquad 
R[\theta] = e^{i\,\theta\,\sigma_2}= \left(\begin{array}{rr}\cos\theta&\sin\theta\\
-\sin\theta&\cos\theta\end{array}\right)\,.
\ea
This allows us to write $\psi_{\omega \vec k}(z)=S[\theta]\,\psi_{\omega k}(z)$ where 
\be\label{eq:rotation.matrices.spinor}
S[\theta]=e^{i\,\theta\,\Sigma_{\ux\uy}}=\cos\frac\theta2 + \Gamma_{\ux\uy}\,\sin\frac\theta2 = 
\left(\begin{array}{rr}
\cos\frac\theta2\,1_{\mbox{\tiny $2\!\!\times\!\! 2$\normalsize}}&-\sin\frac\theta2\,1_{\mbox{\tiny $2\!\!\times\!\! 2$\normalsize}}\\\sin\frac\theta2\,1_{\mbox{\tiny $2\!\!\times\!\! 2$\normalsize}}&\cos\frac\theta2\,1_{\mbox{\tiny $2\!\!\times\!\! 2$\normalsize}}
\end{array}\right)\,,
\ee
is the spinor representation of the rotation matrix (\ref{eq:rotation.matrices.vector}), where $1_{\mbox{\tiny $2\!\!\times\!\! 2$\normalsize}}$ is the $2\times 2$ identity matrix, and we used the explicit form of the generator 
$\Sigma_{{\ux}{\uy}}\equiv \frac{1}{4\,i}\,[\Gamma_{\ux}\,;\Gamma_{\uy}]= 
\frac{1}{2\,i}\,\Gamma_{\ux\uy}$.  The resulting form of the equation (\ref{eq:dirac.equation}) for $\psi_{\omega k}(z)$ now reads
\ba\label{eq:dirac.equation.separated}
\left(\frac1{\sqrt {g(z)}}\,\Gamma_{\uz}\,\partial_z -
\frac i{\sqrt {f(z)}}\,\Gamma_{\ut}\,\left(\omega+ q\,{h(z)}\right)
+ i\,k\,z\,\Gamma_{\ux}- m\,L\right)\;\psi_{\omega k}(z)=0\,.
\ea
Notice that this equation is real due to the choice of gamma matrices (\ref{eq:gamma.matrices}).  

Let us now introduce the projectors 
\be
\label{eq:projectors}
\Pi_\alpha\equiv\frac12
\left(
1+(-1)^{\alpha+1}\; \Gamma^{\uz}\,\Gamma^{\ut}\,\Gamma^{\underline{x}}\right)
=\left\{\begin{array}{l}
\left(\begin{array}{cc}1_{\mbox{\tiny $2\!\!\times\!\! 2$\normalsize}}&0\\0&0\end{array}\right)\qquad,\qquad\alpha=1\\~\\
\left(\begin{array}{cc}0&0\\0&1_{\mbox{\tiny $2\!\!\times\!\! 2$\normalsize}}\end{array}\right)\qquad,\qquad\alpha=2
\end{array}\right.
\ee
With their help we can decompose the Dirac field into ``spin'' states as $\; \psi_{\omega k}(z) = (\Pi_1 + \Pi_2)\,\psi_{\omega k}(z) = 
\sum_{\alpha=1}^2\,\psi_{\alpha\omega k}(z)$, 
where the projected fields are
\be\label{eq:fermion.projected}
\psi_{\alpha\omega k}(z)\equiv\Pi_\alpha \psi_{\omega k}(z)
=\left\{\begin{array}{l}
\left(\begin{array}{c} \psi^{(1)}_{\omega k}(z)\\0\end{array}\right)
\qquad,\qquad \alpha=1\\~\\
\left(\begin{array}{c}0\\\psi^{(2)}_{\omega k}(z)\end{array}\right)
\qquad,\qquad \alpha=2
\end{array}\right.
\ee
where each of the $\psi_{\omega k}^{(\alpha)}$ have two components.
Inserting the resulting decomposition in (\ref{eq:fermion.decomposition.first}), we obtain a solution with energy $\omega$ momentum $\vec k$ and ``spin" $\alpha$ of the form 
\be\label{eq:fermion.decomposition.second}
\Psi_{\alpha\omega\vec k}(t,\vec x,z)=\Psi_{\alpha\omega\vec k}(\vec x,z)e^{-i\omega t}=\frac1{{\cal N}_{\alpha\omega\vec k}}\;
\frac{z}{f(z)^\frac{1}{4}}\,e^{i(\vec k\cdot \vec x-\omega t)}\, S[\theta]\;
\psi_{\alpha\omega k}(z)\,.
\ee
Plugging this back into the field equation (\ref{eq:dirac.equation.separated}) we get a decoupled system for  the bi-spinors $\psi^{(\alpha)}_{\omega k}(z)$, as
\be\label{eq:dirac.equation.separated.spin}
\psi^{(\alpha)'}_{\omega k}(z) + \sqrt{g(z)}\,
\left(\frac{\omega + q\,{h(z)}}{\sqrt{f(z)}}\, i\,\sigma_2 + (-)^\alpha\,k\,z\,\sigma_1 - m\,L\,\sigma_3\right)\,
\psi^{(\alpha)}_{\omega k}(z)= 0\qquad,\qquad\,\alpha=1,2
\ee
{
This system can now be turned into a second order, Schr\" odinger-like equation for a single function $\phi_{\omega k}(z)$, 
that is completely determined (up to normalization) by imposing smoothness at the boundary and smoolthness/ingoing boundary conditions at the horizon. 
We do this in the next subsection
}
\footnote{
It is worth to notice that from the representation (\ref{eq:gamma.matrices}) the generators of the Lorentz subgroup in $2+1$ dimensions are
\be\label{eq:lorentz.generators}
\Sigma_{{\ut}{\ux}}=\frac{1}{2\,i}\,\left(\begin{array}{cc}-\sigma_3&0\\0&\sigma_3\end{array}\right)\quad,\quad
\Sigma_{\ut\uy}=\frac{1}{2\,i}\,\left(\begin{array}{cc}0&\sigma_3\\\sigma_3&0
\end{array}\right)\quad,\quad
\Sigma_{\ux\uy}=\frac{1}{2\,i}\,\left(\begin{array}{cc}0&-1_{\mbox{\tiny $2\!\!\times\!\! 2$\normalsize}}\\1_{\mbox{\tiny $2\!\!\times\!\! 2$\normalsize}}&0\end{array}\right)\,,
\ee
From here it should be clear that $\psi^{(\alpha)}_{\omega k}(z)$ are {not} Dirac spinors in $2+1$ dimensions, since they mix under Lorentz transformations.
}.

\newpage
\subsubsection{The effective Schr\"odinger equation.}  
\label{app:free.fermionic.states.effective.schroedinger}
We want to transform \eqref{eq:dirac.equation.separated.spin} into a second order Schr\"odinger equation to which we can apply our intuitions regarding wave functions. In order to do that, let us introduce the functions 
\be\label{eq:auxiliary.f.plus.minus}
f_k^\pm (z) \equiv \sqrt{g(z)}\;\left( z\,k\pm \frac{\omega + q\,{h(z)}}{\sqrt{f(z)}}\right)\,.
\ee
Now, if we consider first $\alpha=2$, we can parameterize the bi-spinor $\psi^{(2)}_{\omega k}(z)$ in terms of two functions $\psi^{(2)-}_{\omega k}(z)$ and $\phi^{(2)}_{\omega k}(z)$, in the form
\be\label{eq:bispinor.2}
\psi^{(2)}_{\omega k}(z)\equiv \left(\begin{array}{c}\sqrt{f_k^+ (z)}\,\phi^{(2)}_{\omega,k}(z)\\ 
	\psi^{(2)-}_{\omega k}(z)\end{array}\right)\,,
\ee
then substituting in (\ref{eq:dirac.equation.separated.spin}) we get from its components the pair of coupled first order equations
\bea\label{eq:dirac.equation.separated.bispinor.2}
\phi^{(2)}_{\omega k}{}'(z) +\left( \frac{f_k^+{}'(z)}{2\,f_k^+(z)} - m\,L\;\sqrt{g(z)}\right)\;\phi^{(2)}_{\omega k}(z) + \sqrt{f^+_k(z)}\;\psi^{(2)-}_{\omega k}(z) &=& 0\,,\\
\psi^{(2)-}_{\omega k}{}'(z) + 
m\,L\;\sqrt{g(z)}\;\psi^{(2)-}_{\omega k}(z) + f^-_k(z)\sqrt{f^+_k(z)}\;\phi^{(2)}_{\omega k}(z) &=& 0\,,
\eea
from the first of which we can rewrite
\be\label{eq:dirac.equation.solution.bispinor.2}
\psi^{(2)-}_{\omega k}(z) = -\frac{1}{\sqrt{f^+_k(z)}}\;
\left(\phi^{(2)}_{\omega k}{}'(z) +\left(\frac{f_k^+{}'(z)}{2\,f_k^+(z)} - m\,L\;\sqrt{g(z)}\right)\;\phi^{(2)}_{\omega k}(z)\right)\,,
\ee
and now plugging this into the second equation, we get a second order Schr\"odinger-like equation for $\phi^{(2)}_{\omega k}$
\be\label{eq:effective.schroedinger.bispinor.2}
-\phi^{(2)}_{\omega k}{}''(z) + U(z)\;\phi^{(2)}_{\omega k}(z) = 0\,,
\ee
where the potential is given by
\small
\bea\label{eq:potential}
U(z) \!&=&\! g(z)\!\left( k^2\! z^2 \!-\!  \frac{(\omega+q{  h(z)})^2}{f(z)} \!+\!m^2\!L^2\right) 
\!-\! \frac{f_k^+{}''(z)}{2 f_k^+(z)} + \frac{3}{4} 
\left(\frac{f_k^+{}'(z)}{f_k^+(z)}\right)^2\!\!
-mL \sqrt{g(z)}\left(\ln\left(\frac{f_k^+(z)}{\sqrt{g(z)}}\right)\right)'\,,\cr&&
\eea
\normalsize
If instead we consider $\alpha=1$, we  can parametrize the bi-spinor $\psi^{(1)}_{\omega k}(z)$ in terms of the functions $\psi^{(1)+}_{\omega k}(z)$ and $\phi^{(1)}_{\omega k}(z)$, as
\be\label{eq:bispinor.1}
\psi^{(1)}_{\omega k}(z)\equiv \left(\begin{array}{c}
	\psi^{(1)+}_{\omega k}(z)\\\sqrt{f_k^+ (z)}\,\phi^{(1)}_{\omega,k}(z)
\end{array}\right)\,.
\ee
After introducing it in (\ref{eq:dirac.equation.separated.spin}) we get as before a pair of coupled first order equations
\bea\label{eq:dirac.equation.separated.bispinor.1}
\phi^{(1)}_{\omega k}{}'(z) +\left( \frac{f_k^+{}'(z)}{2\,f_k^+(z)} +  m\,L\;\sqrt{g(z)}\right)\;\phi^{(1)}_{\omega k}(z) - \sqrt{f^+_k(z)}\;\psi^{(1)+}_{\omega k}(z) &=& 0\,,\\
\psi^{(1)+}_{\omega k}{}'(z) -
m\,L\;\sqrt{g(z)}\;\psi^{(1)+}_{\omega k}(z) -  f^-_k(z)\sqrt{f^+_k(z)}\;\phi^{(1)}_{\omega k}(z) &=& 0\,.
\eea
From the first equation we can obtain
\be\label{eq:dirac.equation.solution.bispinor.1}
\psi^{(1)+}_{\omega k}(z) = \frac{1}{\sqrt{f^+_k(z)}}\;
\left(\phi^{(1)}_{\omega k}{}'(z) +\left(\frac{f_k^+{}'(z)}{2\,f_k^+(z)} + m\,L\;\sqrt{g(z)}\right)\;\phi^{(1)}_{\omega k}(z)\right)\,,
\ee
and by plugging it into the second equation of (\ref{eq:dirac.equation.separated.bispinor.1}) we get a second order Schr\"odinger-like equation for the function $\phi^{(1)}_{\omega k}$, with the form
\be\label{eq:effective.schroedinger.bispinor.1}
-\phi^{(1)}_{\omega k}{}''(z) + U(z)\;\phi^{(1)}_{\omega k}(z) = 0\,,
\ee
with exactly the same potential (\ref{eq:potential}) as before.

{
We look for solutions of \eqref{eq:effective.schroedinger.bispinor.2} or equivalently \eqref{eq:effective.schroedinger.bispinor.1} which do no diverge neither at the boundary nor at the horizon. In other words, we need normalizable solutions to the Schr\"odinger-like problem}.

A last remark: we could try a solution of equations (\ref{eq:dirac.equation.separated.spin}) such that: 
$\;\psi^{(1)}_{\omega k}(z) = \psi^{(2)}_{\omega (-k)}(z)$ in \eqref{eq:bispinor.2} or  \eqref{eq:bispinor.1}. However, that would not be a smart choice since it would introduce a vanishing denominator at \eqref{eq:dirac.equation.solution.bispinor.2} or \eqref{eq:dirac.equation.solution.bispinor.1} whenever $f^+_k(z)$ vanishes at negative value of the momenta. A better definition is to write (\ref{eq:bispinor.2}) and (\ref{eq:bispinor.1}) with $\phi^{(1)}_{\omega k}(z)=  \phi^{(2)}_{\omega k}(z)\equiv \phi_{\omega k}(z)$ the unique solution of \eqref{eq:effective.schroedinger.bispinor.1}, 
as we will show nextly. 
\subsection{General form of the free fermionic bases}
\label{app:free.fermionic.states.general.form}
\subsubsection{Quantization of the frequencies}
\label{app:free.fermionic.states.general.form.quantization}
We must now impose boundary conditions on our solutions, in order to ensure smoothness everywhere. In particular, we impose regular boundary conditions in the ultraviolet, and we can chose either regular or in-going boundary conditions in the infrared.

The boundary conditions have an important consequence: the frequency $\omega$ is not free but is related to the modulus $k$ of the two-momentum through a dispersion relation that is in general labelled 
by some integer index $m$, resulting in  $\omega=\omega_{m}(k)$ (see section \ref{app:WKB.electron.star} for explicit forms of dispersion relations).  This allows us to replace the indices $k \omega$ in our functions of the previous sections by $km$.

Taking into account these facts, and collecting the results given in  (\ref{eq:dirac.equation.separated.spin})-(\ref{eq:effective.schroedinger.bispinor.1}) we can write for the fermionic mode in (\ref{eq:fermion.decomposition.second}), the general form
\bea\label{eq:fermion.decomposition.final}
\Psi_{\alpha m \vec k}(\vec x,z) 
&=& \frac1{{\cal N}_{\alpha m k}}\;
\frac{z}{f(z)^\frac{1}{4}}\,e^{i \vec k\cdot\vec x}\, e^{i\,\theta\, \Sigma_{\ux\uy}}
\;
\left(\begin{array}{l}
\frac{1-(-1)^{\alpha}}{2}
\psi^{(1)+}_{m k}(z)\\~\\
\frac{1-(-1)^{\alpha}}{2}\sqrt{f_k^+(z)}\;\phi_{mk}(z)\\~\\
\frac{1+(-1)^{\alpha}}{2}\sqrt{f_{k}^+(z)}\;\phi_{mk}(z)\\~\\
\frac{1+(-1)^{\alpha}}{2}\psi^{(2)-}_{m k}(z)
\end{array}\right)\,,\qquad 
\eea 
where $\psi^{(2)-}_{m k}(z)$ and $\psi^{(1)+}_{m k}(z)$ 
are given in (\ref{eq:dirac.equation.solution.bispinor.2}) and (\ref{eq:dirac.equation.solution.bispinor.1}) respectively, in terms of the solution $\phi_{m k}(z)$ of (\ref{eq:effective.schroedinger.bispinor.2}) (or equivalently of (\ref{eq:effective.schroedinger.bispinor.1})). Notice that the $4$-tuple in parenthesis on the right corresponds to what was called $\psi_{\alpha \omega k}(z)$ in equation \eqref{eq:fermion.decomposition.second}, and that will be referred bellow as $\psi_{\alpha m k}(z)$ in attention to the quantization of the frequencies.

\subsubsection{Normalization}
\label{app:free.fermionic.states.general.form.normalization}
To completely define the modes (\ref{eq:fermion.decomposition.final}) we need to fix their normalization. 
To this end we introduce in the space of spinors the following scalar product \cite{Birrell:1982ix}
\be\label{eq:internal.product}
(\Psi_1; \Psi_2)(t) \equiv \int_{\Sigma_t}dz\, d^2 x\;\sqrt{|H|}\;\Psi_1(t,\vec x, z)^\dagger\; \Psi_2(t,\vec x, z)\,,
\ee
where $\Sigma_t$ is the space of constant $t$ and 
\be
H= L^{2}\,\left( g(z)\;dz^2+\frac{d\vec x^2}{z^2}\right)\qquad;\qquad \sqrt{|H|}= L^3\,\frac{\sqrt{g(z)}}{z^2}\,,
\label{eq:spatial.metric}
\ee
is the induced metric on it. 
It is not difficult to prove that in the space of solutions to the Dirac equation (\ref{eq:dirac.equation}) the operator $i\partial_t$ is hermitian, {\em i.e.} 
\be
(\Psi_1; i\partial_t\Psi_2)(t)=(i\partial_t\Psi_1; \Psi_2)(t)\,.
\label{eq:product.hermitean}
\ee
Standard arguments then show that eigenspinors of $i\partial_t$ 
with different eigenvalues are orthogonal with respect to  (\ref{eq:internal.product}). 
By applying this result to the spinors (\ref{eq:fermion.decomposition.final}) 
we know that they result orthogonal for different $m$'s. 
By using this fact we straightforwardly get the orthonormality relation
\be\label{eq:orthogonality}
\left(\Psi_{ \alpha m \vec {k}}; \Psi_{\alpha' m' \vec k'}\right) \equiv \int_{\Sigma_t}dz\, d^2\vec x\;\sqrt{H}\;\Psi^\dagger_{\alpha m \vec {k}}(\vec x,z)\;\Psi_{\alpha' m'\vec k'}(\vec x,z) = \delta_{\alpha\alpha'}\;\delta_{mm'}\; 
\delta^2(\vec k - \vec k')  \,,
\ee
if the normalization constant is fixed such that
\be\label{eq:norm}
|{\cal N}_{\alpha m k}|^2 =  (2\,\pi)^2\,L^3\,\int_0^\infty\,dz\,\sqrt{\frac{g(z)}{f(z)}}\;
\psi_{\alpha m k}(z)^\dagger\;\psi_{\alpha m k}(z)\,.
\ee

Finally, after the calculations above, the general solution of (\ref{eq:dirac.equation}) can be written in coordinates space as
\be\label{eq:fermion.decomposition.annihiltion.operator}
\Psi(t,\vec x,z) = \int\,d^2\vec k\;\sum_{\alpha m}\,c_{\alpha m \vec k}(t)\;\Psi_{\alpha m \vec k}(\vec x,z) \,.
\ee
For later use, it is worth to remind that (\ref{eq:orthogonality}) yields for the above solutions the completeness relation
\be\label{eq:completeness}
\int\,d^2\vec k\;\sum_{\alpha m}\,\Psi_{\alpha m \vec k}(\vec x,z)\;
 \Psi^\dagger_{\alpha m \vec k}(\vec x',z')  = \frac{1}{\sqrt{|H|}}\,\delta(z-z')\;\delta^2(\vec x-\vec x')  \,.
\ee

As we will see in the next section, the above defined normalization provides canonical anti-commutation relations for the corresponding Hamiltonian variables. This allows us to define canonical creation and annihilation operators for the fermionic modes in the bulk. With this, we will be able to establish a standard Landau description for the bulk Fermi liquid.
\newpage
\section{Hamiltonian theory}
\label{app:hamiltonian.theory}
In this Appendix we develop the Hamiltonian theory for the fermions in the bulk, using the orthonormal basis obtained in the previous section. In section \ref{app:hamiltonian.theory.dynamical.setup}, we define our dynamics for the fermionic degrees of freedom, and write the generic form of the corresponding Hamiltonian. Then in section \ref{app:hamiltonian.theory.free} we deal with its free part, while the interacting part is worked out in section \ref{app:hamiltonian.theory.interaction}.
\subsection{Dynamical setup}
\label{app:hamiltonian.theory.dynamical.setup}
The action for our spinor field that propagates in the asymptotically AdS bulk reads
\be
S^\Psi= \int d^4x \sqrt{-|G|}\;{\cal L}^\Psi 
=
-\int d^4x \sqrt{-|G|}\;
\left(
 \bar\Psi\,(\slashed {\cal D}-m)\Psi 
+
 T^{\sigma_1\sigma_2}_{\sigma_3\sigma_4}\; \bar\Psi_{\sigma_1}\;\bar\Psi_{\sigma_2}\;\Psi^{\sigma_3}\;\Psi^{\sigma_4} 
\right)\,,
\label{eq:action.complete}
\ee
where the $\sigma$'s are spin indices running from $1$ to $4$, and we have defined the conjugate spinor $\bar\Psi\equiv \Psi^\dagger\,i\,\Gamma^{\ut}$. We included a four-fermion interaction term, which takes the most general form that respects covariance in four-dimensional curved space and fermion number conservation. It is written in terms of an invariant tensor 
$T^{\sigma_1\sigma_2}_{\sigma_3\sigma_4}$ given by
\ba\label{eq:interaction.tensor.space}
T^{\sigma_1\sigma_2}_{\sigma_3\sigma_4}
&=&  g_1\;\delta^{\sigma_1}{}_{\sigma_3}\;\delta^{\sigma_2}{}_{\sigma_4}
+ g_2\;(\Gamma^5)^{\sigma_1}{}_{\sigma_3}\;(\Gamma^5)^{\sigma_2}{}_{\sigma_4}
+g_3\;(\Gamma^a)^{\sigma_1}{}_{\sigma_3}\;(\Gamma_a)^{\sigma_2}{}_{\sigma_4}\cr
&+&
 g_4\;  
(\Gamma^a\,\Gamma^5)^{\sigma_1}{}_{\sigma_3}\;
(\Gamma_a\,\Gamma^5)^{\sigma_2}{}_{\sigma_4}
-\frac{g_5}{4}\; 
\left([\Gamma^a,\Gamma^b]\,\Gamma^5\right)^{\sigma_1}{}_{\sigma_3}\;
\left([\Gamma_a,\Gamma_b]\,\Gamma^5\right)^{\sigma_2}{}_{\sigma_4}\,,
\ea
where we defined $\;\Gamma^5\equiv-i\,\Gamma^{\ut}\,\Gamma^{\ux}\,\Gamma^{\uy}\,\Gamma^{\uz}\,\,$. 
This tensor satisfies the condition $T^{\sigma_1\sigma_2}_{\sigma_3\sigma_4}=T^{\sigma_2\sigma_1}_{\sigma_4\sigma_3}$, as well as the Lorentz invariance property
\bea
 T^{\sigma_1\sigma_2}_{\sigma_3\sigma_4} 
=
S[\Lambda]^{\sigma_1}{}_{\delta_1}\; 
S[\Lambda]^{\sigma_2}{}_{\delta_2}\;
T^{\delta_1\delta_2}_{\delta_3\delta_4}\;
{S[\Lambda]^{-1}}^{\delta_3}{}_{\sigma_3}\;
{S[\Lambda]^{-1}}^{\delta_4}{}_{\sigma_4}\,,
\label{eq:interaction.tensor.invariance}
\ea
for any Lorentz transformation $\Lambda$, where $S[\Lambda]$ denotes its spinorial representation. This property will be useful in what follows, in particular when we apply it to the rotation on the $\ux\uy$ plane represented by the unitary matrix $S[\theta]$  defined in \eqref{eq:rotation.matrices.spinor}.

From the action \eqref{eq:action.complete} the momentum conjugate to the spinor field is 
\be
\pi=i\,\sqrt{|H|}\,\Psi^\dagger\,,
\label{eq:conjugate.momentum}
\ee
allowing us to write the Hamiltonian as
\be 
H =\int d^2\vec x\,dz\, \left(\pi\,\partial_t\Psi-\sqrt{-|G|}\,{\cal L}^\Psi \right)\,,
\label{eq:hamiltonian.definition}
\ee 
where ${\cal L}^\Psi$ can be read from \eqref{eq:action.complete}.

Inserting the momentum \eqref{eq:conjugate.momentum} into the canonical anti-commutation relations for coordinates and momenta, we get that the spinor field satisfies
\be\label{eq:canonica.commutation.space}
\{\Psi(t,\vec x,z),  \Psi^\dagger(t,\vec x',z')\}  = \frac{1}{\sqrt{|H|}}\,\delta(z-z')\;\delta^2(\vec x-\vec x')  \,.
\ee
Then from the decomposition (\ref{eq:fermion.decomposition.annihiltion.operator}) we get the standard anti-commutation relations 
\be\label{eq:canonica.commutation.momentum}
\{c_{\alpha m\vec k}(t), c^\dagger_{\alpha' m'\vec k'}(t)\} = 
\delta_{\alpha\alpha'}\;\delta_{mm'}\;
\delta^2(\vec k - \vec k')\,,
\ee
identifying $c^\dagger_{\alpha m\vec k}$ and $c_{\alpha m\vec k}$ as the creation and annihilation operators respectively for the fermionic modes in the bulk. In the rest of this section we rewrite the bulk dynamics given by  \eqref{eq:hamiltonian.definition} in terms of them.
\subsection{The free Hamiltonian}
\label{app:hamiltonian.theory.free}
Let us first take the free part of the Hamiltonian \eqref{eq:hamiltonian.definition}, which reads
\small
\bea
H_{\sf free} &=& -\!\int  d^2\vec x\int dz\sqrt{|H|}
\left(
i\Psi^\dagger
\left( 
\frac{i}{2}\omega^{BC}{}_t \Sigma_{BC}
- 
i\, q\, h(z)\, \right)
\Psi\right.
-
\left.\sqrt{|G_{tt}|}
 \bar\Psi\,(\Gamma^{\uz} {\cal D}_{\uz}+\Gamma^{\underline{i}}{\cal D}_{\underline{i}}-m)\Psi 
\right)\cr&&
\label{eq:free.hamiltonian.definition}
\eea
\normalsize
now using the Dirac equation we get
\be 
H_{\sf free} = \frac{1}{2}
\int  d^2\vec x\,\int dz\,\sqrt{|H|}
\;
i\Psi^\dagger\partial_t\Psi
+h.c.\,.
\label{eq:free.hamiltonian.on.shell}
\ee 
Then inserting the decomposition \eqref{eq:fermion.decomposition.annihiltion.operator} and using the orthogonality relation \eqref{eq:completeness} we get the free Hamiltonian in terms of creation and annihilation operators, as
\be 
 H_{\sf free}=\sum_{\alpha m}\int d^2\vec k\;\omega_m(k)\; c^\dagger_{\alpha m\vec k}\;c_{\alpha m\vec k}\,.
\ee 
In other words, the bulk degrees of freedom are described by a set of independent fermionic species labelled with an integer index $m$ corresponding to the mode on the $z$ direction. Each species has a dispersion relation given by $\omega_m(k)$, with a two-fold degeneracy given by the spin $\alpha$. 

\subsection{The interaction Hamiltonian}
\label{app:hamiltonian.theory.interaction}
Regarding the interacting part of the Hamiltonian, we have
\ba\label{eq:interaction.hamiltonian.space}
H_{\sf int}=\int dz\,d^2{\vec x}\,\sqrt{|H|}\, 
\;T^{\sigma_1\sigma_2}_{\sigma_3\sigma_4}\; \bar\Psi_{\sigma_1}\;\bar\Psi_{\sigma_2}\;\Psi^{\sigma_3}\;\Psi^{\sigma_4}\,.
\ea
Plugging the decomposition (\ref{eq:fermion.decomposition.annihiltion.operator}) into (\ref{eq:interaction.hamiltonian.space}) we get
\small
\bea\label{eq:interaction.hamiltonian.momentum}
H_{\sf int}\!&=&\!\!\!\!
\sum_{ \underset{m_1m_2m_3m_4}{\mbox{\tiny$\alpha_1 \alpha_2\alpha_3\alpha_4$\normalsize}}}
\!\!\!
\int d^2\!k_1\dots
 d^2\! k_4\,
\delta^{(2)}\!(\vec k_1 \!+\!\vec k_2 \!-\!\vec k_3 \!- \!\vec k_4)\, 
\;t^{\alpha_1 m_1{\vec k}_1;\alpha_2 m_2{\vec k}_2}_
{\alpha_3 m_3{\vec k}_3;\alpha_4 m_4{\vec k}_4}
\; 
c_{\alpha_1 m_1 {\vec k}_1}^\dagger\;c_{\alpha_2 m_2 {\vec k}_2}^\dagger
c_{\alpha_3 m_3 {\vec k}_3}\;c_{\alpha_4 m_4 {\vec k}_4}\,,\nonumber\\
\eea
\normalsize
where the $\vec x$ integral has been performed explicitly giving origin to the momentum conservation $\delta$-function, while the $z$ integral is contained in the definition of the momentum-space interaction tensor 
\bea\label{eq:interaction.tensor.momentum}
t^{\alpha_1 m_1{\vec k}_1;\alpha_2 m_2{\vec k}_2}_
{\alpha_3 m_3{\vec k}_3;\alpha_4 m_4{\vec k}_4}
&=&L^3(2\pi)^2\,
S[\theta_1]^\dagger{}^{\delta^1}{}_{\!\!\sigma^1}\,
S[\theta_2]^\dagger{}^{\delta^2}{}_{\!\!\sigma^2}\,
T^{\sigma_1\sigma_2}_{\sigma_3\sigma_4}\,
S[\theta_3]^{\sigma^3}\!{}_{\!\delta^3}\,
S[\theta_4]^{\sigma^4}\!{}_{\!\delta^4}\,
I^{\alpha_1 m_1{ k}_1;\alpha_2 m_2{  k}_2;\delta_1\delta_2}_
{\alpha_3 m_3{ k}_3;\alpha_4 m_4{  k}_4;\delta_3\delta_4}\,,
\nonumber\\
\ea
in terms of the integrals
\bea
I^{\alpha_1 m_1{  k}_1;\alpha_2 m_2{  k}_2;\delta_1\delta_2}_
{\alpha_3 m_3{  k}_3;\alpha_4 m_4{  k}_4;\delta_3\delta_4}
&=& 
\int dz\;z^2\;\frac{\sqrt{g(z)}}{f(z)}\;
\frac{\bar\psi_{\alpha_1 m_1 k_1;\delta_1}(z)}{{\cal N}_{\alpha_1 m_1 k_1}}\;
\frac{\bar\psi_{\alpha_2 m_2 k_2;\delta_2}(z)}{{\cal N}_{\alpha_2 m_2 k_2}}\;
\frac{\psi_{\alpha_3 m_3 k_3}^{\delta^3}(z)}{{\cal N}_{\alpha_3 m_3 k_3}}\;
\frac{\psi_{\alpha_4 m_4 k_4}^{\delta^4}(z)}{{\cal N}_{\alpha_4 m_4 k_4}}\,,\cr
&&
\label{eq:fermion.integrals.all}
\ea
where functions $\psi_{\alpha \omega k}(z)$ are defined as in equation \eqref{eq:fermion.decomposition.second} and correspond to the $4$-tuples in parenthesis on the right of equation \eqref{eq:fermion.decomposition.final}. As we show below, only a subset of the integrals \eqref{eq:fermion.integrals.all} is needed for a Landau description of the bulk fluid.
\section{Landau description of the bulk fermions}
\label{app:Landau.description}
In the previous section we rewrite the dynamics of the Dirac spinor in the bulk as that of a bunch of fermionic species in one less dimension, indexed by the energy mode $m$ and the spin $\alpha$. In the present appendix, we develop a description in terms of the Landau theory for a multi-component Fermi liquid. We do that by using a perturbative approach. 
\subsection{Perturbative derivation}
\label{app:Landau.description.derivation}
We work in the grand canonical ensemble at chemical potential $\mu$ and temperature $T=1/\beta$ (that we will take to zero at the end of the calculations). We define the expectation value of an operator ${\cal O}$ by
\be  \label{eq:expectation.value.exact}
\langle {\cal O}\rangle =
\frac{1}{Z}\,{\rm tr}\left(e^{-\beta \left(H - \mu\,N \right)}{\cal O}\right)\,,
\qquad\quad {\rm where}\qquad\quad
Z={\rm tr }\left(e^{-\beta \left(H - \mu\,N \right) }\right) =e^{-\beta\,\Omega(T,\mu )} \,,
\ee
where $N$  is the number of particles and $\Omega(T, \mu)$ is the grand canonical potential. The last formula can be inverted according to
\be
\Omega(T,\mu)=-\frac 1\beta \log Z\,.
\label{eq:grand.canonical.potential.formal}
\ee 

To perform a perturbative approximation, we write the Hamiltonian as
\be
H=H_{\sf free}+H_{\sf int} \,,
\label{eq:hamiltonian.complete}
\ee
and assume that the constants $g_i$ in \eqref{eq:interaction.tensor.space} are small. With the help of the free Hamiltonian $H_{\sf free}$ we define the zeroth order quantities
\be  
\langle {\cal O}\rangle_{\sf free} =
\frac{1}{Z_{\sf free}}\,{\rm tr}\left(e^{-\beta \left(H_{\sf free} - \mu\,N \right)}{\cal O}\right)
\qquad\quad{\rm where}\qquad\quad
Z_{\sf free}={\rm tr }\left(e^{-\beta \left(H_{\sf free} - \mu\,N \right) }\right)  \,.
\label{eq:expectation.value.free}
\ee
Then expanding \eqref{eq:expectation.value.exact} to first order in $g_i$, we obtain
\be
Z\approx Z_{\sf free}\left(1-\beta\langle H_{\sf int}\rangle_{\sf free}\right) \,,
\label{eq:partition.function.perturbed}
\ee
which implies for the grand canonical potential
\be
\Omega(T,\mu)\approx -\frac 1\beta \log Z_{\sf free}
+
\langle H_{\sf int}\rangle_{\sf free}\,.
\label{eq:grand.canonical.potential.perturbed}
\ee 
For the first term in \eqref{eq:grand.canonical.potential.perturbed} we can write
\be
Z_{\sf free}=e^{-\beta \int d^2\vec k\,\sum_{\alpha m  } \;\omega_{m}( k)N_{\alpha m\vec k }}\,,
\label{eq:partition.function.free}
\ee
where $N_{\alpha m\vec k }$ are the occupation numbers of the one-particle states $N_{\alpha m\vec k }= \langle c^\dagger_{\alpha m\vec k }c_{\alpha m\vec k }\rangle $. This implies for the free part of the grand canonical potential
\be\label{eq:grand.canonical.potential.free}
-\frac 1\beta \log Z_{\sf free}= \int d^2\vec k\,\sum_{\alpha m  } \;\omega_{m}( k)N_{\alpha m\vec k }\,.
\ee
On the other hand, to write the second term in \eqref{eq:grand.canonical.potential.perturbed}, we use \eqref{eq:interaction.hamiltonian.momentum} to have
\scriptsize
\be\label{eq:interaction.energy.free}
\langle H_{\sf int}\rangle_{\sf free}=
\!\!\!\!
\sum_{ \underset{m_1m_2m_3m_4}{\mbox{\tiny$\alpha_1 \alpha_2\alpha_3\alpha_4$\normalsize}}}
\!\!\!
\int d^2\!k_1\dots
 d^2\! k_4\,
\delta^{(2)}\!(\vec k_1 \!+\!\vec k_2 \!-\!\vec k_3 \!- \!\vec k_4)\, 
\;t^{\alpha_1 m_1{\vec k}_1;\alpha_2 m_2{\vec k}_2}_
{\alpha_3 m_3{\vec k}_3;\alpha_4 m_4{\vec k}_4}
\,
\left\langle\!
c_{\alpha_1 m_1 {\vec k}_1}^\dagger\,c_{\alpha_2 m_2 {\vec k}_2}^\dagger
c_{\alpha_3 m_3 {\vec k}_3}\,c_{\alpha_4 m_4 {\vec k}_4}\!
\right\rangle_{\sf free}
\ee
\normalsize
Calculating the expectation value in the right hand side 
\scriptsize
\bea
\label{eq:expectation.value.quartic}
&&\left\langle
c_{\alpha_1  m_1  \vec {k}_1}^\dagger
c_{\alpha_2  m_2 \vec {k}_2}^\dagger
c_{ \alpha_3   m_3  \vec k_3}
c_{\alpha_4 m_4  \vec k_4}\right\rangle_{\sf free} 
=N_{\alpha_3 m_3\vec k_3}\,N_{\alpha_4 m_4\vec k_4}\times \\&&\times 
\left(
\delta_{\alpha_1 \alpha_4}\,\delta_{\alpha_2  \alpha_3 }\,
\delta_{m_1 m_4 }\,\delta_{m_2m_3 }\,
\delta^2(\vec {k}_1-\vec k')\,\delta^2(\vec { k}_2-\vec k_3)
-
\delta_{\alpha_1 \alpha_3}\,\delta_{\alpha_2  \alpha_4 }\,
\delta_{m_1 m_3}\,\delta_{m_2 m_4 }\,
\delta^2(\vec {k}_1-\vec k_3)\,\delta^2(\vec { k}_2-\vec k_4) 
\right)\,,
\nonumber
\eea
\normalsize
we replace back in \eqref{eq:interaction.energy.free} to get
\ba\label{eq:interaction.energy.free.2}
\left\langle H_{\sf int}\right\rangle_{\sf free}&=&
\frac12
\int \!d^2 \vec k\,  d^2 \vec k'\,\sum_{\alpha m \alpha' m'}
\,f_{\alpha m \alpha' m'}(\vec k,\vec k')\, 
N_{\alpha m\vec k}\, N_{\alpha' m'\vec k'}\,,
\ea
where the ``Landau interaction functions'' $f_{\alpha m \alpha' m'}(\vec k,\vec k')$ are defined according to
\be\label{eq:interaction.function}
f_{\alpha m \alpha'm'}(\vec k,\vec k') \equiv
t^{\alpha m{\vec k};\alpha' m'{\vec k}'}_
{\alpha' m'{\vec k}';\alpha m{\vec k}} -
t^{\alpha m{\vec k};\alpha' m'{\vec k}'}_
{\alpha m{\vec k};\alpha' m'{\vec k}'} +
t^{\alpha' m'{\vec k}';\alpha m{\vec k}}_
{\alpha m{\vec k};\alpha' m'{\vec k}'} -
t^{\alpha' m'{\vec k}';\alpha m{\vec k}}_
{\alpha' m'{\vec k}';\alpha m{\vec k}}\,.
\ee
Here we used the definition \eqref{eq:interaction.tensor.momentum}, and re-absorbed into the couplings an infinite constant $\delta^2(\vec 0)$ coming from collapsing the momentum integrals\footnote{Formally this is accomplished by regularizing through the introduction of a finite volume in coordinate space, and then taking the infinite volume limit at the end of the calculations.}. The Landau interaction functions contain all the information about the interactions on the multi-component Fermi liquid. 
They verify the following relations
\be
f_{\alpha m \alpha'm'}(\vec k,\vec k') = 
f_{\alpha' m' \alpha m}(\vec k',\vec k) \qquad;\qquad 
f_{\alpha m \alpha m}(\vec k,\vec k)=0\,.
\ee

{
At this point of the discussion, a remark is in order. 
It is usual for condense matter theorists to think about the Landau  interaction function as a correlation function. 
More precisely, let us introduce the index $A\equiv (\alpha m \vec k)$ 
and denote by
\be\label{eq: 1PI}
\Gamma^{(1PI)}_{A_1A_2; A_3A_4}
\equiv \left\langle c_{A_1}\,c_{A_2}\,c^\dagger_{A_3}\,c^\dagger_{A_4}\right\rangle_{1PI}
\ee
the one-particle irreducible vertex that describes the interacction of two elementary excitations.   
Then it can be shown that
\footnote{{
As they are not essential for the argument, we do not care here with numerical factors and propagators residues, that in any case are irrelevant  in  the order at which we work, see \cite{nozieresdupuis} for details.}
}
\be\label{eq:intfunct}
f_{A A'} \sim \Gamma^{(1PI)}_{A A'; A' A}
\ee
At leading order in perturbation theory the right hand side of \eqref{eq: 1PI} reads
\cite{peskin}
\bea\label{eq:gammalo}
\left.\Gamma^{(1PI)}_{A_1A_2; A_3A_4}\right|_{{\sf leading ~order}}
&=& - i\,\left\langle\int dt\, H_{\sf int}\; c_{A_1}\,c_{A_2}\,c^\dagger_{A_3}\,c^\dagger_{A_4}\right\rangle\cr
&\sim& t^{A_1 A_2}_{A_4 A_3}- t^{A_1 A_2}_{A_3 A_4} + 
t^{A_2 A_1}_{A_3 A_4}- t^{A_2 A_1}_{A_4 A_3}
\eea
where in the first line all the operators are in the interaction picture, with the interacting Hamiltonian given in  \eqref{eq:interaction.hamiltonian.momentum}, and in the
second line we  omitted a momentum conservation delta-function. 
By plugging \eqref{eq:gammalo} into \eqref{eq:intfunct} 
we get our result \eqref{eq:interaction.function}. 
}

Now we can use \eqref{eq:interaction.energy.free.2} and \eqref{eq:interaction.function} to write the grand canonical potential as
\be \label{eq:grand.canonical.potential.perturbed.final}
\Omega(T,\mu)=
\int d^2\vec k\,\sum_{\alpha m  } \;\omega_{m}( k)N_{\alpha m\vec k }
+
\frac12
\int \!d^2 \vec k\,  d^2 \vec k'\,\sum_{\alpha m \alpha' m'}
\,f_{\alpha m \alpha' m'}(\vec k,\vec k')\, 
N_{\alpha m\vec k}\, N_{\alpha' m'\vec k'}\,.
\ee 

A perturbation $\delta N_{\alpha m\vec k }$ of the ground state occupation numbers $N_{\alpha m\vec k }$ gives us a variation of the grand canonical potential with the form
\be \label{eq:Landau.formula.app}
\delta \Omega(T,\mu)=
\int d^2\vec k\,\sum_{\alpha m  } \;\epsilon_{m}( k)\delta N_{\alpha m\vec k }
+
\frac12
\int \!d^2 \vec k\,  d^2 \vec k'\,\sum_{\alpha m \alpha' m'}
\,f_{\alpha m \alpha' m'}(\vec k,\vec k')\, 
\delta N_{\alpha m\vec k}\, \delta N_{\alpha' m'\vec k'}\,,
\ee 
where the quasiparticle dispersion relation $\epsilon_m(k)$ has been defined as
\be
\label{eq:dispersion.relation.app}
\epsilon_m(k)=\omega_m(k)+\sum_{ \alpha' m'}\int \!d^2 \vec k'\,   f_{\alpha m \alpha' m'}(\vec k,\vec k')\, N_{\alpha' m'\vec k'}\,.
\ee
For any perturbation  $\delta N_{\alpha m\vec k }$ the quantity $\delta \Omega(T,\mu)$ must be positive to have a stable ground state. Whenever it becomes negative, an instability is triggered.

\subsection{Explicit form of the Landau interaction functions}
\label{sec:Landau.description.explicit.form}
We want to obtain a more explicit form of the Landau interaction functions $f_{\alpha m\,\alpha' m'}(\vec k, \vec k')\;$ suitable of being implemented into a numerical code. The integrals \eqref{eq:fermion.integrals.all} we need are
\bea
I^{\alpha m{k};\delta_3\delta_4}_{\alpha' m'{k}';\delta_1\delta_2}
&=&
I^{\alpha m {\vec k};\alpha' m'{\vec k}';\delta_1\delta_2}_
{\alpha' m'{\vec k}';\alpha m{\vec k};\delta_3\delta_4}\,,
\label{eq:fermion.integrals.restricted}
\eea
and they satify $I^{\alpha m{k};\delta_3\delta_4}_{\alpha' m'{k}';\delta_1\delta_2}=I^{\alpha' m'{k}';\delta_4\delta_3}_{\alpha m{k};\delta_2\delta_1}$ what can be used to rewrite (\ref{eq:interaction.tensor.momentum}) as 
\bea
t^{\alpha m{\vec k};\alpha' m'{\vec k}'}_
{\alpha' m'{\vec k}';\alpha m{\vec k}}
&=&
S[\theta-\theta']^\dagger{}^{\delta^1}{}_{\sigma_1}\;
 T^{\sigma_1\delta_2}_{\delta_3\sigma_2}\;
S[\theta-\theta']^{\sigma^2}{}_{\delta^4}\;
I^{\alpha m{k};\delta_3\delta_4}_{\alpha' m'{k}';\delta_1\delta_2}
\label{eq:interaction.tensor.momentum.integrals.restricted.1}\,,
\\
t^{\alpha m{\vec k};\alpha' m'{\vec k}'}_
{\alpha m{\vec k};\alpha' m'{\vec k}'}
&=&
S[\theta-\theta']^\dagger{}^{\delta^1}{}_{\sigma_1}\;
T^{\sigma_1\delta_2}_{\sigma_2\delta_3}\;
S[\theta-\theta']^{\sigma^2}{}_{\delta^4}\;
I^{\alpha m{k};\delta_3\delta_4}_{\alpha' m'{k}';\delta_1\delta_2}\,.
\label{eq:interaction.tensor.momentum.integrals.restricted.2}
\eea
%
%
To further disentangle the angle dependence, we find convenient to use the explicit form of the spinorial rotations matrices (\ref{eq:rotation.matrices.spinor}) to write
\be
\label{eq:rotation.matrices.auxiliary.tensor.P}
S[\theta-\theta']^\dagger{}^{\delta_1}{}_{\sigma_1}\;
S[\theta-\theta']^{\sigma^2}{}_{\delta^4}=
P_0{}^{\delta_1\sigma_2}_{\sigma_1\delta_4} +
\cos(\theta-\theta')\;P_c{}^{\delta_1\sigma_2}_{\sigma_1\delta_4} + 
\sin(\theta-\theta')\;P_s{}^{\delta_1\sigma_2}_{\sigma_1\delta_4}\,,
\ee
where the constant, cosine and sine auxiliary tensors are given respectively as 
\bea
\label{eq:auxiliary.tensor.P}
P_0{}^{\delta_1\sigma_2}_{\sigma_1\delta_4}&\equiv& 
\frac12\;\left(\delta^{\delta_1}{}_{\sigma_1}\;\delta^{\sigma_2}{}_{\delta_4} - 
\Gamma_{\ux\uy}{}^{\delta_1}{}_{\sigma_1}\;\Gamma_{\ux\uy}{}^{\sigma_2}{}_{\delta_4}\right)=+P_0{}^{\sigma_2\delta_1}_{\delta_4\sigma_1}\,,\cr
P_c{}^{\delta_1\sigma_2}_{\sigma_1\delta_4}&\equiv&
\frac12\;\left(\delta^{\delta_1}{}_{\sigma_1}\;\delta^{\sigma_2}{}_{\delta_4} +
\Gamma_{\ux\uy}{}^{\delta_1}{}_{\sigma_1}\;\Gamma_{\ux\uy}{}^{\sigma_2}{}_{\delta_4}
\right)=+P_c{}^{\sigma_2\delta_1}_{\delta_4\sigma_1}\,,\cr
P_s{}^{\delta_1\sigma_2}_{\sigma_1\delta_4}&\equiv&
\frac12\;\left(\delta^{\delta_1}{}_{\sigma_1}\;
\Gamma_{\ux\uy}{}^{\sigma_2}{}_{\delta_4} - 
\Gamma_{\ux\uy}{}^{\delta_1}{}_{\sigma_1}\;\delta^{\sigma_2}{}_{\delta_4}\right)
=-P_s{}^{\sigma_2\delta_1}_{\delta_4\sigma_1}\,.
\eea
%
%
By inserting this expressions into \eqref{eq:interaction.tensor.momentum.integrals.restricted.1}-\eqref{eq:interaction.tensor.momentum.integrals.restricted.2} we finally obtain from \eqref{eq:interaction.function} a completely factorized form for the Landau interaction functions, as
\be\label{eq:Landau.parameters}
 f_{\alpha m;\alpha' m'}(\vec k, {\vec k}') = 
 f_{\alpha m k;\alpha' m' k'}^0 + 
\cos(\theta-\theta')\;f_{\alpha m k;\alpha' m' k'}^c + 
\sin(\theta-\theta')\;f_{\alpha m k;\alpha' m'{ k}'}^s\,,
\ee
written in terms of three independent constant, cosine and sine components, which are given  according to
\bea\label{eq:cuulo}
f_{\alpha m{k};\alpha' m'{k}'}^0&=& 
I^{\alpha m{k};\delta_3\delta_4}_{\alpha' m'{k}';\delta_1\delta_2}\;
\left(
 T^{\delta_1\delta_2}_{\delta_3\delta_4} -  T^{\delta_1\delta_2}_{\delta_4\delta_3} 
+ \Gamma_{\ux\uy}{}^{\delta_1}{}_{\sigma_1}\, 
\left( T^{\sigma_1\delta_2}_{\sigma_4\delta_3} -  T^{\sigma_1\delta_2}_{\delta_3\sigma_4} \right)\,
\Gamma_{\ux\uy}{}^{\sigma_4}{}_{\delta_4}\right)\cr
&&\cr
&=&+f_{\alpha' m'{k}';\alpha m{k}}^0\,,\cr
&&\cr
 f_{\alpha m{k};\alpha' m'{k}'}^c&=& 
I^{\alpha m{k};\delta_3\delta_4}_{\alpha' m'{k}';\delta_1\delta_2}\;
\left(
 T^{\delta_1\delta_2}_{\delta_3\delta_4} -  T^{\delta_1\delta_2}_{\delta_4\delta_3} 
- \Gamma_{\ux\uy}{}^{\delta_1}{}_{\sigma_1}\, 
\left( T^{\sigma_1\delta_2}_{\sigma_4\delta_3} -  T^{\sigma_1\delta_2}_{\delta_3\sigma_4} \right)\,
\Gamma_{\ux\uy}{}^{\sigma_4}{}_{\delta_4}\right)\cr
&&\cr
&=&+f^c_{\alpha' m'{k}';\alpha m{k}}\,,\cr
&&\cr
 f_{\alpha m{k};\alpha' m'{k}'}^s&=& 
I^{\alpha m{k};\delta_3\delta_4}_{\alpha' m'{k}';\delta_1\delta_2}\;
\left(
\left( T^{\delta_1\delta_2}_{\delta_3\sigma} - 
 T^{\delta_1\delta_2}_{\sigma\delta_3}\right)\,
\Gamma_{\ux\uy}{}^{\sigma}{}_{\delta_4}\,
- \Gamma_{\ux\uy}{}^{\delta_1}{}_{\sigma}\,
\left( T^{\sigma\delta_2}_{\delta_3\delta_4}-
 T^{\sigma\delta_2}_{\delta_4\delta_3}\right)
\right)\cr
&&\cr
&=&-f_{\alpha' m'{k}';\alpha m{k}}^s\,.
\eea
Notice that we dropped any auxiliary quantity, having obtained a formula for the Landau interaction functions which is completely written in terms of the interaction tensor \eqref{eq:interaction.tensor.space} and the integrals \eqref{eq:fermion.integrals.restricted} of the form \eqref{eq:fermion.integrals.all}, where the radial functions were defined in equations \eqref{eq:fermion.decomposition.second}-\eqref{eq:fermion.decomposition.final} in terms of the solutions of equation \eqref{eq:effective.schroedinger.bispinor.1}.
\newpage
\section{Pomeranchuk method}
\label{app:Pomeranchuk.method}
We give in this section a brief introduction to Pomeranchuk's method to detect instabilities in a Landau fermi liquid. 
We focus on the isotropic, two dimensional case, since it is what concern us in this paper.
We begin by explaining the method for a single spinless fermion (section  \ref{app:Pomeranchuk.method.simple}), the generalization for multiple species of spinful fermions is given later (section \ref{app:Pomeranchuk.method.general}).
\subsection{Single spinless fermion}
\label{app:Pomeranchuk.method.simple}
Given a two-dimensional system of spinless fermion with quasiparticle dispersion relation $\epsilon(k)$, its Fermi surface is defined in momentum space by the relation $\epsilon(k_F)=0$. 
In the ground state all the single quasiparticle states with momenta $k\leq k_F$ are occupied, while those with momenta $k>k_F$ are empty. In other words, we can write the occupation number as $N_{\vec k}= {\sf H}(-\epsilon(k))$, where ${\sf H}$ is the Heavyside step function.

A fermionic excitation can then be represented by a small deformation of the Fermi surface, charactherized by a variation of the occupation numbers $\delta N_{\vec k}$. The excitation energy at weak coupling is then  given by Landau's formula (\ref{eq:Landau.formula.app})
\be
\delta \Omega = 
\int\; d^2\vec k\;\epsilon(k)\;\delta N_{\vec k}+ 
\frac12\int \!\!d^2\vec k\; d^2\vec k'\; f(\vec k,\vec k')\; 
\delta N_{\vec k}\;\delta N_{\vec k'}\,,
\label{eq:Landau.formula.simple} 
\ee
where $f(\vec k,\vec k')$ is the Landau interaction function. The variation $\delta N_{\vec k}$ on the occupation numbers of the state take the values $\delta N_{\vec k}=0,\pm 1$ at any point with $\vec k$ of momentum space. They can be parametrized as
\be
\delta N_{\vec k}={\sf H}\big{(}\!-\!\epsilon(k) + \delta g(\vec k)\big{)} - {\sf H}\left(\!- \epsilon(k) \right)\,,
\label{eq:occupation.number.app}
\ee
where $\delta g(\vec k)$ is an auxiliary function that characterizes the deformation of the Fermi surface. 
By using the expansion of the Heavyside function in terms of Dirac delta function and its derivatives, we get\footnote{This is a shorthand calculation, if the reader is uncomfortable with it, (\ref{eq:occupation.number.explanded}) can be smoothed by making the replacement ${\sf H}(x)\rightarrow 1/(1+\exp(-b^2\, x))$ and then taking the $b^2\to\infty$ limit.}
\be\label{eq:occupation.number.explanded}
\delta N_{\vec k}=\delta\big{(}\!-\!\epsilon(k) \big{)}\; \delta g(\vec k) 
+\frac12\;\delta'\big{(}\!-\epsilon(k) \big{)}\;\delta g(\vec k)^2  +\dots\,.
\ee
When replacing (\ref{eq:occupation.number.explanded}) into \eqref{eq:Landau.formula.simple}, the Dirac delta functions enforce the integrands to be evaluated at $\epsilon(k)=0$, namely at the Fermi surface. 
Using polar coordinates in momentum space, this implies that the integrand must be  evaluated at $k=k_F$ and then the momentum integrals reduce to angular ones
\be\label{eq:Landau.formula.angles}
\delta \Omega \!= \frac{k_F}{v_F}\; 
\int \!d\theta \,\frac12\,\delta g(\theta)^2  
+ \frac{k_F{}^2}{v_F{}^2}\;\int \!d\theta\; d\theta'\;\frac12\, f(\theta-\theta')\;\delta g(\theta)\;\delta g(\theta')\,,
\ee
where by using rotational invariance we wrote 
$\;f(\theta-\theta')=f(k_F,\theta;k_F,\theta')\;$ 
and $\;\delta g(\theta)=\delta g(k_F, \theta)$. Here the Fermi velocity is defined as $v_F=d \epsilon(k)/dk|_{k=k_F}$. 

A Fourier expansion now allows us to write
\be\label{eq:interaction.function.angles}
f(\theta-\theta')=   \sum_{n=0}^\infty f_n^c\;\cos\left(n\,(\theta-\theta')\right) + 
\sum_{n=1}^\infty f_n^s\;\sin\left(n\,(\theta-\theta')\right)\,,
\ee
and similarly, to parameterize the deformations by 
the amplitudes $\{ \delta g_n^c;\delta g_n^s\}$ in the decomposition
\be\label{eq:delta.g.modes}
\delta g(\theta)= 
\sum_{n=0}^\infty \delta g_n^c\;\cos\left(n\,\theta\right) + 
\sum_{n=1}^\infty \delta g_n^s\;\sin\left(n\,\theta\right)\,.
\ee
Replacing (\ref{eq:interaction.function.angles}) and (\ref{eq:delta.g.modes}) in (\ref{eq:Landau.formula.angles}) we obtain: 
\be\label{eq:Landau.formula.modes}
\delta \Omega = 
\pi\,\frac{k_F}{v_F}\;\left(1+2\,\pi\,\frac{k_F}{v_F}\;f^c_0\right)\;\delta {g^c_0}^2
+\frac{\pi}{2}\,\frac{k_F}{v_F}\;
\sum_{n=1}^\infty\,\left(1+\pi\,\frac{k_F}{v_F}\; f_n^c\right)\;
\left(\delta g_n^c{}^2+\delta g_n^s{}^2\right)\,.
\ee
We see that the excitation energy ends up written as a quadratic form in the deformation amplitudes $\{\delta g_n^c;\delta g_n^s\}$. 
In order to have a stable system, $\delta \Omega$ has to be positive for any possible excitation, or in other words for any possible amplitudes. 
This implies that the above defined quadratic form must be positive definite, 
leading to the stability conditions:
\ba\label{eq:Pomeranchuk.conditions.simple}
1+2\,\pi\,\frac{k_F}{v_F}\;f^c_0&>&0\,,\cr
1+\pi\,\frac{k_F}{v_F}\; f_n^c&>&0\,,\qquad\qquad\forall\, n\in\mathbb{N}\,.
\ea
If any of these conditions is violated, the system becomes unstable. 
Notice that the parameters $f_n^s$ completely disappear from the calculation. 
The parameters $f_n^c$ are called the ``Landau parameters'' of the Fermi liquid. 

~

It is important to stress that, when working in perturbation theory, the dispersion relation is defined by \eqref{eq:dispersion.relation.app}  as
\be
\label{eq:dispertion.relation.simple}
\epsilon(k)=\omega(k)+\int \!d^2 \vec k'\,   f(\vec k,\vec k')\, N_{\vec k'}\,.
\ee
This implies that only the zeroth order forms of the Fermi momentum and Fermi velocity have to be kept in equations \eqref{eq:Pomeranchuk.conditions.simple}. Indeed, since the second term in those inequalities contains a Landau parameter $f^c_n$, it is already first order in the coupling constants. Thus we can replace \eqref{eq:Pomeranchuk.conditions.simple} by
\ba\label{eq:Pomeranchuk.conditions.simple.free}
1+2\,\pi\,\frac{k_F^{\sf free}}{v_F^{\sf free}}\;f^c_0&>&0\,,\cr
1+\pi\,\frac{k_F^{\sf free}}{v_F^{\sf free}}\; f_n^c&>&0\,,\qquad\qquad\forall\, n\in\mathbb{N}
\ea
where $k_F^{\sf free}$ is obtained from $\omega(k_F^{\sf free})=0$ and we defined $v_F^{\sf free}=d\omega(k)/dk|_{k=k_F^{\sf free}}$.

\subsection{Multiple fermionic species with spin}
\label{app:Pomeranchuk.method.general}
The above procedure can be easily generalized to many species of spinful fermions, the resulting quadratic form being in general non-diagonal in the indices denoting spin $\alpha$ and species $m$. 
As we know, the excitation energy of such fermionic system is given by  (\ref{eq:Landau.formula.app})
\be\label{eq:Landau.formula.general}
\delta \Omega =
\sum_{\alpha m}\int \!\!d^2k\, \epsilon_{m}(k)\; \delta N_{\alpha m}(\vec k)
+\frac12\sum_{\alpha m \alpha'  m' }\int \!\!d^2k\; d^2k'\;
f_{\alpha m \alpha' m'}(\vec k,\vec {k}')\;
\delta N_{\alpha m}(\vec k)\;\delta N_{\alpha'm'}(\vec {k}')\,.
\ee
When the deformations of the occupation numbers get spin and species indices 
$\delta N_{\alpha m}(\vec k)$, so do the functions $\delta g_{\alpha m}(\vec k)$ parameterizing the deformations
\be
\delta N_{\alpha m}(\vec k)={\sf H}\big{(}\!-\!\epsilon_{ m}(k) + \delta g_{\alpha m}(\vec k)\big{)} - {\sf  H}\left(\!- \epsilon_{ m}(k) \right)\,.
\label{eq:ocupation.numbers.general}
\ee
Notice that each species and spin component has its own Fermi momentum $k_F^{ m}$ at which it dispersion relation vanishes $\epsilon_{ m}(k_F^{ m})=0$. 
Going through the same steps as before we get for the energy fluctuation
\be\label{eq:Landau.formula.angles.general}
\delta  \Omega \!= \sum_{\alpha m}\,
\frac{k_F^{ m}}{v_F^{ m}}
\int \!d\theta \;\frac12\,\delta g_{\alpha m}(\theta)^2
+\frac12\!\!  \sum_{\alpha m '\alpha  m'}
\frac{k_F^{ m}}{v_F^{ m}}\,
\frac{k_F^{' m'}}{v_F^{' m'}}\!
\int \!d\theta\, d\theta'\,
f_{\alpha m \alpha'  m'}(\theta-\theta')\,
\delta g_{\alpha m}(\theta)\;\delta g_{\alpha' m'}(\theta'),
\ee
where 
$\;f_{\alpha m \alpha'm'}(\theta\!-\!\theta')= f_{\alpha m \alpha'm'}(k_F^{\alpha m},\theta;\,k_F^{\alpha'm'},\theta')$ and $v_ F^{ m}=d\epsilon_{ m}(k)/dk|_{k=k_F^{ m}}$.

Now we must decompose the interaction function as in \eqref{eq:interaction.function.angles}
\be\label{eq:interaction.function.angles.general}
f_{\alpha m \alpha'm'}(\theta-\theta')= \sum_{n=0}^\infty {f_{\alpha m \alpha'm'}}_n^c\;\cos\left(n\,(\theta-\theta')\right) + 
\sum_{n=1}^\infty {f_{\alpha m \alpha'm'}}_n^s\;\sin\left(n\,(\theta-\theta')\right)\,.
\ee
Although from (\ref{eq:Landau.parameters}) we can already see that in our case only the modes $n=0,1$ are present in the decomposition \eqref{eq:interaction.function.angles.general}, we mantain the general analysis for now. The deformation parameters can be decomposed in parallel with (\ref{eq:delta.g.modes}) to get 
\be\label{eq:delta.g.general}
\delta g_{\alpha m}(\theta)= 
\sum_{n=0}^\infty {\delta g_{\alpha m}}_n^c\;\cos\left(n\,\theta\right) + 
\sum_{n=1}^\infty {\delta g_{\alpha m}}_n^s\;\sin\left(n\,\theta\right)\,,
\ee
in terms of the deformation amplitudes $\{{\delta g_{\alpha m}}_n^c,{\delta g_{\alpha m}}_n^s\}$.

Going ahead to plug the decompositions \eqref{eq:interaction.function.angles.general} and \eqref{eq:delta.g.general} into \eqref{eq:Landau.formula.angles.general}, we get the generalization of (\ref{eq:Landau.formula.modes}) to be
\ba\label{eq:Landau.formula.modes.general}
\delta \Omega\!&=& \!
\sum_{\alpha m \alpha'm' }\,\pi\,\frac{k_F^{ m}}{v_F^{ m}}\;
\left(\delta_{\alpha m \alpha' m'}+2\,\pi\,
\frac{k_F^{ m'}}{v_F^{ m'}}\;
{f_{\alpha m \alpha' m'}}^c_0\right)\;
\delta {g_{\alpha m}}^c_0\;{\delta g_{\alpha' m'}}^c_0+\cr
&&+ 
\sum_{n=1}^\infty\,\sum_{\alpha m \alpha'm' }
\frac{\pi}2\,\frac{k_F^{  m}}{v_F^{  m}}\;
\left(\delta_{\alpha m \alpha' m' }+\pi
\frac{k_F^{ m'}}{v_F^{ m'}}
{f_{\alpha m \alpha' m' }}_n^c\right)
\left( {\delta g_{\alpha m}}_n^c\;
{\delta g_{\alpha' m'}}_n^c +
{\delta g_{\alpha m}}_n^s\;{\delta g_{\alpha' m'}}_n^s\right)+\cr
&&+
\sum_{n=1}^\infty\,\sum_{\alpha m \alpha' m'}\;
\frac{\pi^2}2\,\frac{k_F^{  m}}{v_F^{  m}}\;
\frac{k_F^{ m'}}{v_F^{ m'}}\;
{f_{\alpha m \alpha' m' }}_n^s\;
\left({\delta g_{\alpha m}}_n^s\;{\delta g_{\alpha' m'}}_n^c-{\delta g_{\alpha m}}_n^c\;{\delta g_{\alpha' m'}}_n^s\right)\,.
\ea
\normalsize
On a stable state, this quadratic form has to be positive definite. 
Neccesary and sufficient conditions for stability can be obtained by considering the following two disantangled cases. 
\begin{itemize}
\item Putting ${\delta g_{\alpha m}}^c_0\neq 0$ and $\;{\delta g_{\alpha m}}_n^c=\delta {g_{\alpha m}}_n^s=0$ for $n\geq1$, we have that the first line  in (\ref{eq:Landau.formula.modes.general}) must be positive definite, that applying Sylvester's criterion is equivalent to, 
\ba\label{eq:Pomeranchuk.conditon.0}
\left|\frac{k_F^{  m}}{v_F^{  m}}\;
\left(\delta_{\alpha m \bar\alpha \bar m} +
2\,\pi\,\frac{k_F^{ m'}}{v_F^{m'}}\;
{f_{\alpha m\alpha'm' }}^c_0\right)
\right|_{M\times M}>0\;&,&\;\forall M\in \mathbb{N}\,,
\ea
where $|\cdots|_{M\times M}$ stands for the $M$-th minor, that is the determinant of the $M\times M$ upper-left submatrix. 

\item Considering now the opposite case:
${\delta g_{\alpha m}}^c_0 = 0$ and ${\delta g_{\alpha m}}_n^c\neq 0, \delta {g_{\alpha m}}_n^s\neq 0$ for $n\geq1$,
we have to impose the positivity of the quadratic form:  $\delta\Omega =\sum_{n=1}^{\infty}\delta\Omega_n$ defined by the second and third lines in (\ref{eq:Landau.formula.modes.general}) for each $n$ separetely,  since they do not couple. 
To simplify the analysis, we find convenient to introduce the perturbation vectors in indices $(\alpha m)$:  
$\vec u_n\equiv(\delta g_{\alpha m}{}_n^c)$ and 
$\vec v_n\equiv(\delta g_{\alpha m}{}_n^s)$. 
Then we can write,
\be\label{eq:deltaOmegan}
\delta\Omega_n = \vec u_n{}^t\;{\bf S_n}\;\vec u_n 
+ \vec v_n{}^t\;{\bf S_n}\;\vec v_n 
- \vec u_n{}^t\;{\bf A_n}\;\vec v_n
+ \vec v_n{}^t\;{\bf A_n}\;\vec u_n\,,
\ee 
where the matrices ${\bf S_n}={\bf S_n}{}^t$ and 
${\bf A_n}=-{\bf A_n}{}^t$ are read from (\ref{eq:Landau.formula.modes.general}). 
By further defining the complex vectors 
$\vec z_n\equiv \vec u_n + i\;\vec v_n\;$, (\ref{eq:deltaOmegan}) takes the simple form 
\be
\delta\Omega_n =
\vec z_n{}^\dagger\;{\bf H_n}\;\vec z_n\qquad;\qquad 
{\bf H_n} = {\bf S_n} + i\;{\bf A_n}\,.
\ee
From here it is clear that the positivity condition of  
$\delta\Omega$ is just that the hermitian matrices  
${\bf H_n}={\bf H_n}{}^\dagger$ must be positive, 
for any $n\in\mathbb{N}$.  
By using the explicit expressions of ${\bf S_n}$ and 
${\bf A_n}$, this is equivalent to ask that
\be\label{eq:Pomeranchuk.conditon.sc}
\left|\frac{k_F^{  m}}{v_F^{  m}}\;
\left(\delta_{\alpha m \alpha' m' }+\pi
\frac{k_F^{ m'}}{v_F^{ m'}}\,\left({f_{\alpha m \alpha' m' }}_n^c+i{f_{\alpha m \alpha' m' }}_n^s\right) \right)\,
\right|_{M\times M}
>0\quad,\quad\forall M\in \mathbb{N}\,.
\ee

\end{itemize} 
In conclusion, if any of the minors in (\ref{eq:Pomeranchuk.conditon.0}) and (\ref{eq:Pomeranchuk.conditon.sc})
is negative, the quadratic form has a negative mode 
that would lead to an excitation with negative energy: $\,\delta  \Omega<0$, 
thus triggering an instability on the fermionic system.
\subsection{Summary and application to the holographic setup}
\label{sec:summary}
In summary, in order to perform the stability analysis, all we need are 
\begin{itemize}
\item The value of the free Fermi momenta $k_F^{ m\;{\sf free}}$, obtained from $\omega_{m}(k_F^{ m\;{\sf free}})=0$. Indeed, notice that, since $k_F^m$ is always positive, we can remove it from the pre-factor in \eqref{eq:Pomeranchuk.conditon.0} and \eqref{eq:Pomeranchuk.conditon.sc}. Regarding the second term, since $k_F^m$ is multiplied by the interaction function, we can replace it by $k_F^{ m\;{\sf free}}$.
\item The value of the free Fermi velocities $v_F^{ m\;{\sf free}}=\left.d\omega_m(k)/dk\right|_{k=k_F^{m\;{\sf free}}}$, as long as we assume that $v_F^m$ is always positive in other to remove it from the overall pre-factor in \eqref{eq:Pomeranchuk.conditon.0} and \eqref{eq:Pomeranchuk.conditon.sc}.  
\item The Fourier components of the Landau parameters, evaluated at the free Fermi momenta $k_F^{ m\;{\sf free}}$. 
Notice that equation \eqref{eq:Landau.parameters} implies that only the modes with $n=0, 1$ enter into the stability analysis in the holographic setup. The Landau parameters are obtained from \eqref{eq:cuulo} in terms of the integrals \eqref{eq:fermion.integrals.restricted} evaluated at the free Fermi momenta. Since $\omega_{m}(k_F^{ m\;{\sf free}})=0$, we need only static wave functions in \eqref{eq:fermion.integrals.restricted}. 
\end{itemize} 
\section{The electron star background}
\label{app:electron.star}
In this appendix, we summarize the construction of the electron star solution. We do so in order to fix notation and to have a fully self-contained discussion. However, all the results presented in this section were obtained in the original reference \cite{Hartnoll:2010gu} to which the interest reader is referred. 

\bigskip 

We work with the previously presented Ansatz \eqref{eq:background.detailed} 
\bea\label{eq:electron.star.background}
ds^2 &=& L^2\left(
-f(z)\, dt^2+g(z)\, dz^2 +\frac{d\vec x^2}{z^2}
\right)\,,\cr
A&=& h(z)\,dt\,.
\eea
The field equations of $(3+1)$-dimensional Einstein-Maxwell theory in the presence of a negative cosmological constant $-3/L^2$ plus matter in the form of a charged perfect fluid are
\bea\label{eq:electron.star.equations}
R_{MN}-\frac{1}{2}\,g_{MN}\,R -\frac{3}{L^{2}}\,g_{MN}&=&
\kappa^2\,\left(T^{\sf Maxwell}_{MN}+T^{\sf Fluid}_{MN}\right)\,,\cr 
\nabla_{N}F^{MN}&=&e^{2}\,J^M\,,
\eea
with $\kappa$ and $e^2$ the gravitational and electromagnetic couplings respectively.  
Here the contributions to the energy-momentum tensor and current density read 
\bea
\label{eq:electron.star.energymomentum}
T^{\sf Maxwell}_{MN}&=&\frac{1}{e^{2}}\,\left(F_{MP}\,F_{N}^{P}-
\frac{1}{4}\,g_{MN}\,F_{PQ}\,F^{PQ}\right)\,,\cr 
T^{\sf Fluid}_{MN}&=&(\rho + p)\,u_{M}\,u_{N} + p\,g_{MN}\,,\cr 
J_{M}&=&\sigma\, u_{M}\,,
\eea
The functions $p$, $\rho$ and $\sigma$ are the pressure, energy density and charge density of the fluid, respectively, and $u_M$ its four-velocity. Furthermore, the conservation equations must hold
\be\label{eco}
\nabla^N T_{MN}=0\qquad;\qquad \nabla_M J^M=0\,.
\ee
Since we aim to describe a relativistic system of fermions of mass $m$, 
we can try a description of the energy and charge densities in terms of 
the density of states in flat space of a free fermion gas $\;g(E)=\pi^{-2}\,E\,(E^{2}-m^{2})^{1/2}$. This description would be accurate as long as the wavelength of the fermions is much shorter than the local curvature radius. 
Furthermore, we must consider the equation of state of the system in terms of a grand canonical potential $\Omega = -p\,V$. 
At zero temperature the fluid functions read
\bea\label{eq:sigma.rho.p}
\sigma&=&\int_{m}^{\varepsilon_{F}} dE\, g(E)
=\frac{1}{\pi^2}\,\int_{m}^{\epsilon_{F}}  dE\, E\,(E^{2}-m^{2})^{1/2}\,,\cr 
\rho&=&\int_{m}^{\varepsilon_{F}}dE\,g(E)\,E
=\frac{1}{\pi^2}\,\int_{m}^{\epsilon_{F}} dE\,E^{2}\,(E^{2}-m^{2})^{1/2}\,,\cr
p &=& -\rho + \varepsilon_F\,\sigma\,.
\eea
where $\varepsilon_F$ is the bulk Fermi energy. Consistency with Ansatz (\ref{eq:electron.star.background}) imposes that all the functions 
be dependent only on $z$. Moreover, the velocity of the fluid must be unitary, which identifies it with the vierbein temporal vector, {\em  i.e. } $u^M\,\partial_M \equiv e_{\underline t}= 1/(L\,\sqrt f)$, see (\ref{eq:vielbein}). 
This last fact implies that the bulk Fermi energy coincides with the local ({\em i.e.} measured by comoving observers) chemical potential $\;\varepsilon_F\equiv {h}/L{\sqrt f}$. 

It is convenient to work with the dimensionless quantities 
\be\label{eq:sigma.rho.p.hatted}
\hat p = \kappa^2\,L^2\,p\qquad,\qquad 
\hat\rho= \kappa^2\,L^2\,\rho\qquad,\qquad 
\hat\sigma = e\,\kappa\,L^2\,\sigma\qquad,\qquad
\hat h = h/\gamma\,.
\ee
Here we introduced $\gamma\equiv {e\,L}/{\kappa}$.
In terms of these scaled variables (\ref{eq:sigma.rho.p}) can be integrated to obtain the explicit expressions
\bea\label{eq:sigma.rho.p.explicit}
\hat{\sigma}&=&\frac{\hat{\beta^2}}{3}\,\left(\frac{\hat h^{2}}{f}-\hat m^2\right)^\frac{3}{2}\,,\cr
\hat{\rho}&=&\frac{\hat\beta^2}{8}\,\left[\sqrt{\frac{\hat h^{2}}{f}-\hat{m}^2}\;
\left(-\hat m^2\,\frac{\hat h}{\sqrt f} + 2\,\frac{\hat h^3}{f^\frac{3}{2}}\right) -
\hat m^4\,\ln\left(\frac{\hat h}{\sqrt f} + \sqrt{\frac{\hat h^2}{f}-\hat m^2}\right) + 
\hat m^4\,\ln\hat m\right]\,,\cr
\hat p &=& -\hat \rho + \frac{\hat h}{\sqrt f}\,\hat\sigma\,,
\eea
where we introduced the two independent parameters $\hat \beta$ and $\hat m$ that can be used to define the electron star 
\be\label{eq:beta.m}
\hat{\beta}\equiv\frac{e\,\gamma}{\pi}\qquad,\qquad \hat{m}\equiv\frac{m\,L}{\gamma}\,.
\ee
By plugging (\ref{eq:electron.star.background}) and (\ref{eq:sigma.rho.p.explicit}) in (\ref{eq:electron.star.equations}) we get the field equations  
\bea\label{eq:electron.star.eom}
\frac{z\,f^{\prime}}{f}+\frac{z\,g^{\prime}}{g}
+ \hat\sigma\, z^2\,g\,\frac{\hat h}{\sqrt f} + 4 &=&0\,,\cr
\frac{z\,f'}{f} -\frac{ z^2\,\hat h'^2}{2\,f}+(3+\hat{p})\,z^2\,g- 1&=&0\,,\cr 
\frac{z^2\,\hat h''}{h} + \hat{\sigma}\,z^2\,g\,\frac{\hat h}{\sqrt f}\,  \left(\frac{z\,\hat h'}{2\,\hat h} 
- \frac{f}{\hat h^2}\right)&=& 0\,.
\eea
These equations have to be solved numerically for each value of $\hat m$ and $\hat \beta$ to obtain the metric and gauge functions $(f,g,\hat h)$.

Regarding the boundary conditions in the IR, we observe from  (\ref{eq:sigma.rho.p.explicit}) that the electron star extends from infinity to the 
radius $z_s$ such that 
\be
\label{eq:electron.star.boundary}
\hat p(z_s) = \hat\rho(z_s) = \hat\sigma(z_s)=0\,,\qquad\Longleftrightarrow\qquad
T^{\sf Fluid}_{MN}(z_s)=J_{M}(z_s) =0\,,
\ee
whose position is determined by the equation
\be\label{eq:electron.star.boundary.definition}
\hat h(z_s)^{2} - \hat m^2\;f(z_s) =0\,.
\ee
Outside the electron star, {\em i.e.} $0< z< z_s$, we have that $\sigma=p=\rho=0$, implying that the equations of motion (\ref{eq:electron.star.eom}) reduce to
\be\label{eq:electron.star.eom.outside}
\frac{f^{\prime}}{f}+\frac{g^{\prime}}{g}+\frac{4}{z}=0\qquad;\qquad
\frac{1}{z}\,\frac{f^{\prime}}{f}-\frac{\hat h^{\prime 2}}{2\,f}+3\,g- \frac{1}{z^2}=0\qquad;\qquad
\hat h^{\prime\prime}=0\,,
\ee 
whose general solution is the AdS Reissner-N\"ordstrom black hole
\be\label{eq:Reissner-Nordstrom.solution}
f(z)=\frac{c^2}{z^{2}}-\hat{M}\, z + \frac{\hat{Q}^{2}}{2}\,z^{2}\qquad,\qquad
g(z)=\frac{c^{2}}{z^{4}\,f(z)}\qquad,\qquad 
\hat h(z)=\hat{\mu}-\hat{Q}\,z\,.
\ee
To specify this solution we have to give the four integration constants. 
$(c, \hat M, \hat\mu,\hat Q)$. In the electron star context they must be obtained by matching 
$(f, g,\hat  h, \hat h')$ at the radius  $z=z_s$ with the solution inside the star. 

\bigskip

On the other hand, in the IR limit $z\rightarrow\infty$ the solution acquires the form of a Lifshitz metric. 
In fact, an Ansatz for large $z$ of the form:
\be\label{eq:electron.star.ir.boundary.conditions}
f=\frac{1}{z^{2\lambda}}\,\left(1+f_1\,z^\alpha + \dots\right)\quad,\quad g=\frac{g_\infty}{z^2}\,\left(1+ g_1\,z^\alpha + \dots\right) \quad,\quad \hat h=\frac{h_\infty}{z^\lambda}\,\left(1 + h_1\,z^\alpha + \dots\right)\,,
\ee
solves (\ref{eq:electron.star.eom}) if $\alpha$ takes some of the following three values
\be\label{eq:electron.star.alphas}
\alpha_0=2+\lambda\qquad,\qquad  
\alpha_\pm=1+\frac{2}{\lambda}\pm 
\frac{1}{2}\,\sqrt{
\frac{9\lambda^{3}-21\lambda^{2}+40\lambda -28 -\hat{m}^{2}\lambda(4-3\lambda )^{2}}{(1-\hat{m}^{2})\lambda-1}\,.
}
\ee
The parameter $\lambda$ is called the dynamical critical exponent. 
One can show that an IR non-singular, asymptotically exact Lifshitz solution exists only if 
\be\label{eq:electron.star.lambda.limits}
1\leq\frac{1}{1-\hat m^2}\leq \lambda\,,
\ee
and the root selected is the negative one:  $\alpha=\alpha_-$, 
see \cite{Hartnoll:2010gu} for details.
From the leading order of (\ref{eq:electron.star.eom}) we get, 
\bea\label{eq:electron.star.Lifshitz.parameters}
g^{2}_{\infty}&=&\frac{36\,\lambda^4\,(\lambda-1)}
{\hat \beta^4\,\left(\lambda\,(1-\hat m^2)-1\right)^3}\,,\cr
h^{2}_{\infty}&=&1-\frac{1}{\lambda}\,,\cr 
g_{\infty}\,(3+\hat{p}(\infty))&=& 1+2\,\lambda + \frac{h_\infty{}^2}{2}\,\lambda^{2} \,.
\eea
The first two equations determine the coefficients that define the Lifshitz metric in terms of $\lambda$ and $\hat m$, while that the last equation relates implicitly  $\hat{\beta}$ in terms of $\lambda$ and $\hat{m}$ also. 
So it is possible, and convenient to make numerics, to think the Lifshitz parameter and the mass as the free parameters of the electron star. 
The expansions (\ref{eq:electron.star.ir.boundary.conditions}) result completely determined 
by ($\lambda, \hat m$), and by $f_1$ that remains free. However, the scalings $(t,x, y, z)\rightarrow (b\,t, a\,x,a\,y,a\,z)$ leave the Ansatz (\ref{eq:electron.star.background}) invariant and induce the invariance of the system  (\ref{eq:electron.star.eom}) under
\vspace{-.6cm}
\bea\label{eq:electron.star.scalings}
z\rightarrow a\,z\quad&,&\quad g\rightarrow a^{-2}\,g\,,\cr
f \rightarrow b^{-2}\,f\quad&,&\quad \hat h\rightarrow b^{-1}\,\hat h\,.
\eea
These scalings  allow to fix the leading order term of $f$ in \eqref{eq:electron.star.ir.boundary.conditions}   together with the absolute value of $f_1$, 
while the sign to get the solution with the right behaviour in the UV results to be the negative one.  
In the paper we work out the electron star solution by fixing $f_1=-1$.

\begin{figure}[H]
\begin{center}
\includegraphics[width=4.9cm]{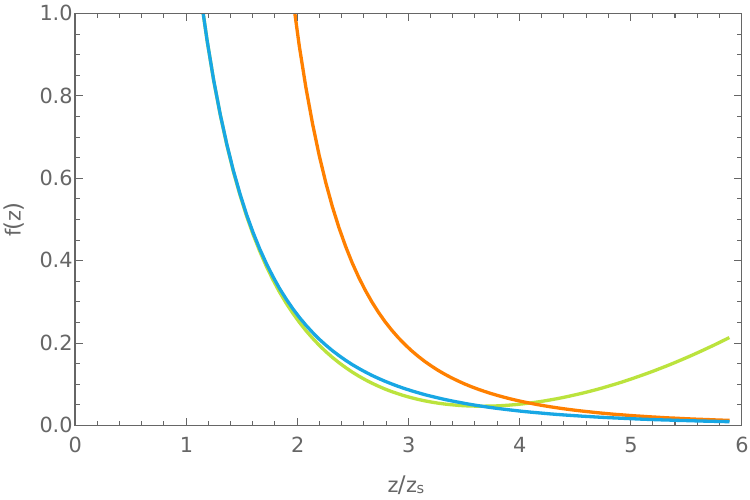}
\includegraphics[width=4.9cm, height=3.18cm]{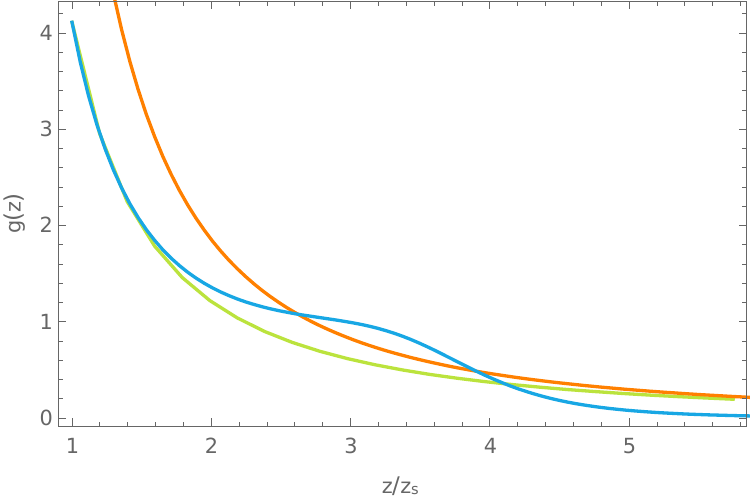}
\includegraphics[width=4.9cm]{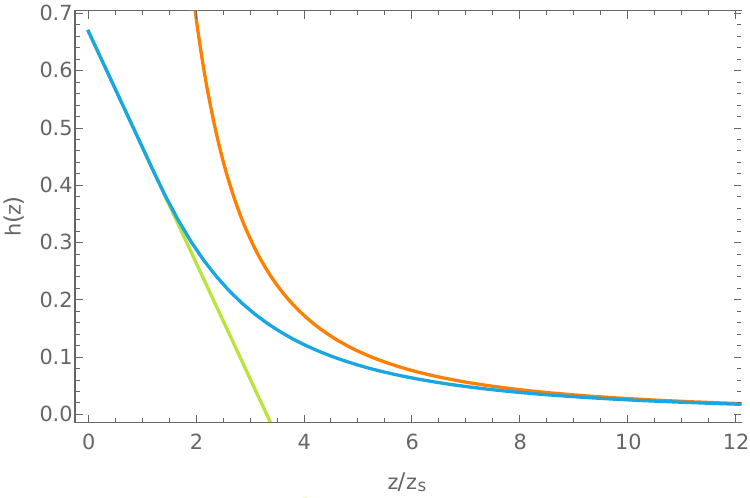}
\caption{  {We plot the electron star background functions 
$f$, $g$ and $h$ vs. $z/z_s$ (light blue line), compared to the Lifshitz solution (orange) and AdS Reissner-N\"ordstrom black hole (lihgt green) respectively. The parameters are chosen as $\hat m=0.4$ and $\lambda=2$.}}
\label{fig:electron.star.profiles}
\end{center}
\end{figure}

In figure \ref{fig:electron.star.profiles} we present the electron star background solution obtained after solving numerically the system (\ref{eq:electron.star.eom}). 
We can see how the electron star solution matches the AdS-RN and Lifshitz solutions in the UV ($z/z_s\ll 1$) and IR ($z/z_s\gg 1$) respectively.   
	 
An important point to be stressed is that, in the weak gravity regime both $\kappa\ll L$ and  
$\kappa\ll 1/m$ conditions should hold, or what is the same, 
$e^2\ll\pi\,\hat\beta$ and $e\ll 1/{\hat m}$. 
On the other hand, in \cite{Hartnoll:2010gu} was shown that the flat treatment (\ref{eq:sigma.rho.p.explicit}) is consistent only if we restrict the electron star parameters to the region $\,\hat\beta\sim 1,\, \hat m\sim 1$. 
Together with the weak gravity conditions, they imply that 
\be\label{fig:electron.star.gamma.condition}
\gamma\sim \frac{1}{e}\sim m\,L \gg 1\,.
\ee 
We will see in appendix \ref{app:WKB} that these conditions are closely linked to the validity of the WKB approximation for the fermionic perturbations.  
\newpage
\section{The WKB solution}
\label{app:WKB}
In the electron star background, we are assuming that we have a large number of particles inside one AdS radius. This is equivalent to the statement that the particle wavelength is much shorter than the characteristic length of the background in which it is moving. This implies that we can solve the effective Schr\"odinger equation \eqref{eq:effective.schroedinger.bispinor.2} (or equivalently \eqref{eq:effective.schroedinger.bispinor.1}) in the Wentzel-Kramers-Brillouin (WKB) approximation, as we do in the rest of this section.
\subsection{Basics of the WKB approximation}
\label{app:WKB.general}
Let us remember basic facts about the WKB approximation (see for example \cite{merzbacher1998quantum}). Let us consider the one-dimensional Schr\"odinger equation written in the form
\be\label{eq:Schroedinger.equation}
-\phi''(z) + U(z)\;\phi(z) = 0\,.
\ee
We will assume that the potential is repulsive and positive at the boundary, implying that at $z$ close enough to $z=0$ we have $U(z)>0$ and $U'(z)<0$. As we move into positive values of $z$ we have a first turning point $z_0$ at which $U(z_0)=0$ and $U'(z_0)<0$. At larger values of $z$ additional turning points $z_r$ appear at which $U(z_r)=0$.

Away from any given turning point, we introduce the functions
\be\label{eq:WKB.waves}
u_\pm (z; z_r) = |U(z)|^{-\frac{1}{4}}\;\exp{\left(\pm \int_{z_r}^z dz\,\sqrt{U(z)}\right)}\,.
\ee
As can be checked by direct substitution, they are good approximate solutions of  \eqref{eq:Schroedinger.equation} as long as $U(z)$ is large enough, which means in particular that we can use them away from the turning points. 
On the other hand, close to any turning point we can linearize the potential, and the solution is then written in terms of Airy functions $A\mbox{\scriptsize $i$\normalsize}$ and $B\mbox{\scriptsize $i$\normalsize}$.
Then the WKB approximate solution around $z_r$ takes the form
\bea\label{eq:WKB.regions}
\phi^{\sf WKB}(z) = \left\{\begin{array}{lcrcl} L^{(r)}_+\; u_+(z; z_r) + L^{(r)}_-\; u_-(z; z_r)\qquad&,&\qquad z_r -z&\gg& |U'(z_r)|^{-\frac{1}{3}}\\~\\
A^{(r)}_+\;A\mbox{\scriptsize $i$\normalsize}(w) + A^{(r)}_-\;B\mbox{\scriptsize $i$\normalsize}(w)\big|_{w=U'(z_r)^{-\frac{1}{3}}\,(z-z_r)}&,&
|z-z_r|&\ll&\left|\frac{2\,U'(z_r)}{U''(z_r)}\right|\\~\\
R^{(r)}_+\; u_+(z; z_r) + R^{(r)}_-\; u_-(z; z_r)\qquad&,&\qquad z-z_r&\gg&|U'(z_r)|^{-\frac{1}{3}}
\end{array}\right.\qquad
\eea
where $L_\pm^{(r)}, R_\pm^{(r)}$ and $A_\pm^{(r)}$ are numerical constants to be determined in order to ensure the boundary conditions and the continuity around each turning point. 

The coefficients of the functions $u_\pm(x,x_r)$ at left and right of a given turning point are linearly related, as in 
\be\label{eq:WKB.matrix}
\left(\begin{array}{c}R^{(r)}_+\\R^{(r)}_-\end{array}\right) = {\bf M}(z_r)\;\left(\begin{array}{c}L^{(r)}_+\\L^{(r)}_-\end{array}\right)\,,
\ee
where the explicit form of the matrix ${\bf M}(z_r)$ is obtained by matching the functions $u_\pm(z;z_r)$ across the turning point using the intermediate Airy form of the solution. It has the expression ${\bf M}(z_r)={\bf M}$ when $U'(z_r)>0$, and ${\bf M}(z_r)={\bf M^{\dagger}}$ when $U'(z_r)<0$, with 
\be\label{eq:WKB.matrix.explicit}
{\bf M} \equiv e^{+i\frac{\pi}{4}}\;\left(\begin{array}{cc}
1&-i\\-\frac{i}{2}&\frac{1}{2}
\end{array}\right)\,.
\ee

The relation between the approximate WKB solutions around any pair of successive turning points can be found by shifting the limits of integration in \eqref{eq:WKB.waves} from $z_r$ to $r_{r+1}$ 
\bea
\label{eq:WKB.shifts}
u_\pm (z;z_r) &=& \varphi_r^\pm\;u_\pm (z;z_{r+1}) \,,
\eea
in terms of the connection coefficients $\varphi_r^\pm$, which read
\bea
\label{WKB.shift.explicit}
\varphi_r^\pm &=& \exp{\left(
\pm  \int_{z_r}^{z_{r+1}} dz\,\sqrt{U(z)}
\right)} \,.
\eea
Compatibility with the form (\ref{eq:WKB.regions}) for the solution around each turning point thus implies the additional linear relation
\be
\label{eq:WKB.matrix.shifts}
\left(\begin{array}{c}L^{(r+1)}_+\\L^{(r+1)}_-\end{array}\right) ={\bf W}(z_r)\;  
\left(\begin{array}{c}R^{(r)}_+\\R^{(r)}_-\end{array}\right) \,,
\ee
in terms of the connection matrix
\be
{\bf W}(z_r)=
\left(\begin{array}{cc}\varphi_r^+ &0\\0&\varphi_r^-\end{array}\right)\,.
\label{eq:WKB.matrix.shitfs.explicit}
\ee

With all the above, we can express the whole set of coefficients of the solution around each of the turning points, in terms of $L^{(0)}_+$ and $L^{(0)}_-$, as follows
\bea\label{eq:WKB.matrix.complete}
\left(\begin{array}{c}L^{(r)}_+\\L^{(r)}_-\end{array}\right) &=& 
{\bf W}(z_{r-1})
\;{\bf M}(z_{r-1})\;\dots
{\bf W}(z_0)\;{\bf M}(z_0)\;
\left(\begin{array}{c}L^{(0)}_+\\L^{(0)}_-\end{array}\right)\,,
 \nonumber\\\nonumber~\\
\left(\begin{array}{c}R^{(r-1)}_+\\R^{(r-1)}_-\end{array}\right) &=& 
{\bf M}(z_{r-1})\; 
{\bf W}(z_{r-2})\;{\bf M}(z_{r-2})\;\dots
{\bf W}(z_{0})
\;{\bf M}(z_0)\;
\left(\begin{array}{c}L^{(0)}_+\\L^{(0)}_-\end{array}\right)\,.
\eea
From this, we can write the relation between the initial (leftmost) and final (rightmost) coefficients for a system with a total of $R$ turning points. It takes the form
\be\label{eq:WKB.matrix.V}
\left(\begin{array}{c}R^{(R-1)}_+\\R^{(R-1)}_-\end{array}\right) = 
{\bf V}\; 
\left(\begin{array}{c}L^{(0)}_+\\L^{(0)}_-\end{array}\right)\,,
\ee
where 
\be\label{eq:WKB.matrix.V.explicit}
{\bf V}=
\left(\begin{array}{cc} V_{11} & V_{12}\\ V_{21} & V_{22}\end{array}\right)
=
{\bf M}(z_{R-1})\; 
{\bf W}(z_{R-2})\;{\bf M}(z_{R-2})\;\dots
{\bf W}(z_{0})
\;{\bf M}(z_0) \,.
\ee

In the present context,  Bohr-Sommerfeld  quantization relations arise when we impose boundary conditions at both extremes of the $z$ axis. This leads to constraints on the matrix elements of ${\bf V}$ and consequently on the parameters of the theory.
\subsection{Application to the electron star background}
\label{app:WKB.electron.star}
After the re-scalings $\omega= \gamma\,\hat{\omega},\;k=\gamma\,\hat{k}$, together with \eqref{eq:sigma.rho.p.hatted} and \eqref{eq:beta.m}, the potential \eqref{eq:potential} entering into the effective Schr\"odinger equation acquires the following form
\small
\bea\label{eq:WKB.electron.star.potential}
U(z)&=&\gamma ^2\, g(z)\,\left( \hat{k}^2\, z^2 -  \frac{(\hat{\omega}+ q\,\hat h(z))^2}{f(z)} +\hat{m}^2 \right) + 
\gamma\,\hat{m}\,\sqrt{g(z)}\;\left(\ln\left( f_k^+(z)\,\sqrt{g(z)}\right)\right)'
+\frac{\left(\sqrt{ f_k^+(z)}\right)''}{\sqrt{f_k^+(z)}}.\cr&&
\eea
\normalsize
As we will be working in an electron star  background, 
we have to consider the large $\gamma$ limit. 
From (\ref{eq:WKB.electron.star.potential}) we then see that this regime coincides with the semiclassical limit where the WKB approximation is reliable. In such limit we can neglect the last two terms in (\ref{eq:WKB.electron.star.potential}) and consider 
\be\label{eq:WKB.electron.star.potential.simplified}
U(z)\approx \gamma ^2\, g(z)\,\left( \hat{k}^2\, z^2 -  \frac{(\hat{\omega}+ q\,\hat h(z))^2}{f(z)} +\hat{m}^2 \right)\,.
\ee 

Close to the AdS boundary $z=0$, we can replace the functions $f, g$ and $\hat h$ by their corresponding UV expansions \eqref{eq:Reissner-Nordstrom.solution}, obtaining $U(z)\sim \gamma^2\,\hat m^2/z^2$. This implies that the potential diverges at the boundary. Since the resulting $u_-(z,z_0)$ function in \eqref{eq:WKB.waves} diverges as we move into the boundary $z=0$, in order to have a normalizable solution, we need to impose $L_-^{(0)}=0$.  

In the deep IR, the functions take their Lifshitz form \eqref{eq:electron.star.ir.boundary.conditions}, and we get $U(z)\sim \gamma^2 {g_\infty}\,( \hat{k}^2-  \hat{\omega}^2z^{2(\lambda-1)} )$. In consequence, for zero frequency the potential goes to a positive constant at infinity. The corresponding function $u_+(z,z_{R-1})$ diverging, we need to impose $R^{(R-1)}_+=0$. Combined with the conditions at the boundary we obtain $R^{(R-1)}_+=V_{11}L^{(0)}_+=0$ or in other words
\be
V_{11}=0\,,
\label{eq:Bohr-Sommerfeld.zero.freuency}
\ee
where the matrix element $V_{11}$ was defined in \eqref{eq:WKB.matrix.V.explicit} in terms of integrals involving the potential $U(z)$. 
For finite frequency on the other hand, the potential diverges into negative values. This results in an $u_-(z,z_{R-1})$ with the form of a wave moving into smaller $z$ values, {\em i.e.} entering the bulk from infinity, which implies that we must impose $R^{(R-1)}_-=0$ in order to avoid causality issues. Since $R^{(R-1)}_-=V_{21}L^{(0)}_+=0$ this implies
\be
V_{21}=0\,.
\label{eq:Bohr-Sommerfeld.finite.freuency}
\ee
Equations \eqref{eq:Bohr-Sommerfeld.zero.freuency} and \eqref{eq:Bohr-Sommerfeld.finite.freuency} impose a constraint between the parameters entering into $U(z)$. 
It often has a discrete set of solutions, being then understood as a quantization condition (see below).

According to the summary on section \ref{sec:summary}, using this information we need to obtain the free Fermi momentum $k_F^{m\;{\sf free}}$ and Fermi velocity $v_F^{ m\;{\sf free}}=\left.d\omega_m(k)/dk\right|^{}_{k=k_F^{m\;{\sf free}}}$ for each mode $m$ in the $z$ direction, as well as its static wave function $\psi_{\alpha m{k_F^{m\,{\sf free}}}}^{\delta}(z)$ or in other words the solution $\phi_{k\omega}^{(2)}$ of the effective Schr\" odinger equation \eqref{eq:effective.schroedinger.bispinor.2} with $k=k_F$ and $\omega=0$.
\newpage
\subsubsection{Fermi momenta and static wave functions} 
\label{app:WKB.electrons.star.Fermi.momenta}
Of particular interest to our problem are the Fermi momenta, 
{\em i.e.} the values ${k_F^{m}}^{\sf free}$ of the momentum $k$ satisfying $\hat\omega_m({k_F^{m}}^{\sf free})=0$. 
This implies that the corresponding potential in \eqref{eq:WKB.electron.star.potential.simplified} has a vanishing frequency, and then goes to a positive constant at infinity. In consequence it has none or an even number of turning points. 

The two last terms in the parenthesis in \eqref{eq:WKB.electron.star.potential.simplified} are positive outside the star, and negative inside it, according to the rule  \eqref{eq:electron.star.boundary.definition}. This implies that for $k$ large enough, the parenthesis remains always positive and there are no turning points. On the other hand, for $k$ smaller than some critical value $k^*$, the potential becomes negative somewhere inside the star, giving rise to a pair of turning points $z_0,z_1$. Only in this last case we are able to find values of the free Fermi momenta, all of them inside a ``Fermi ball'' \cite{Lee:2008xf} of radius $k^*$.

\begin{figure}[H]
\begin{center}
\includegraphics[width=7.8cm]{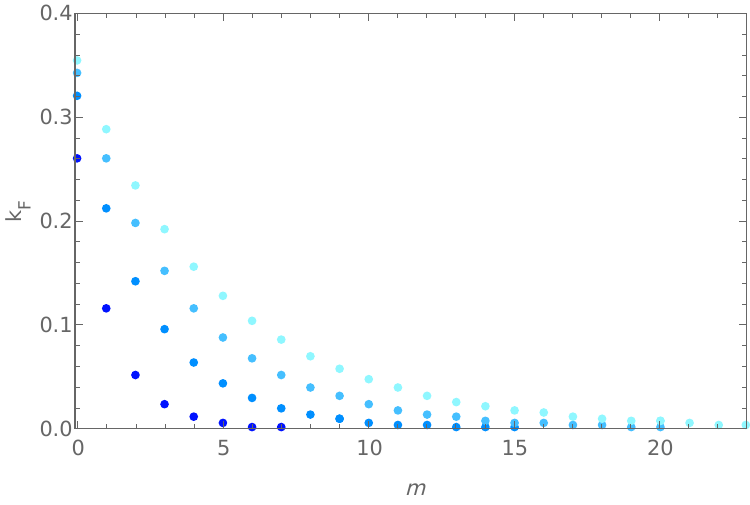}
\vspace{-4mm}
\caption{\label{fig:Fermi.momenta}
The Fermi mom\-enta are plot\-ted in terms of $m$ for dif\-ferent val\-ues of $\gamma=$  $100,$ $150,$ $200,$ $250$ from bot\-tom to top. 
For each value of $\gamma$ there exists a maximum value of the Fermi momentum.
}
\end{center}
\end{figure}

\begin{figure}[H]
\begin{center}
\includegraphics[width=7cm]{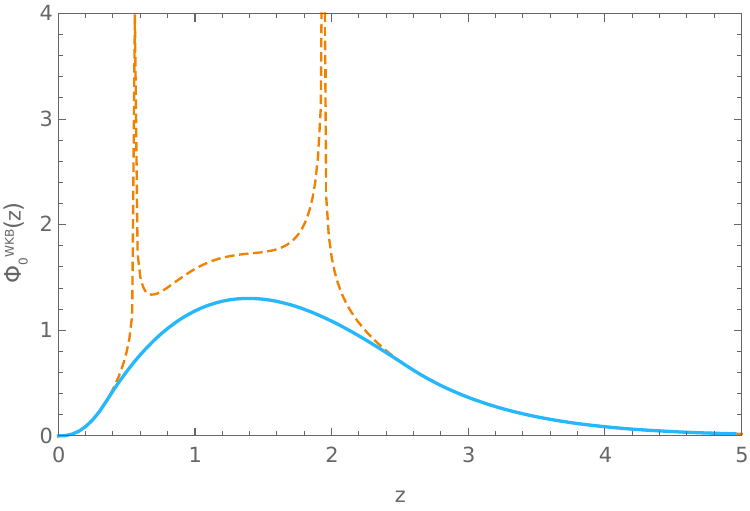}\quad
\includegraphics[width=7cm]{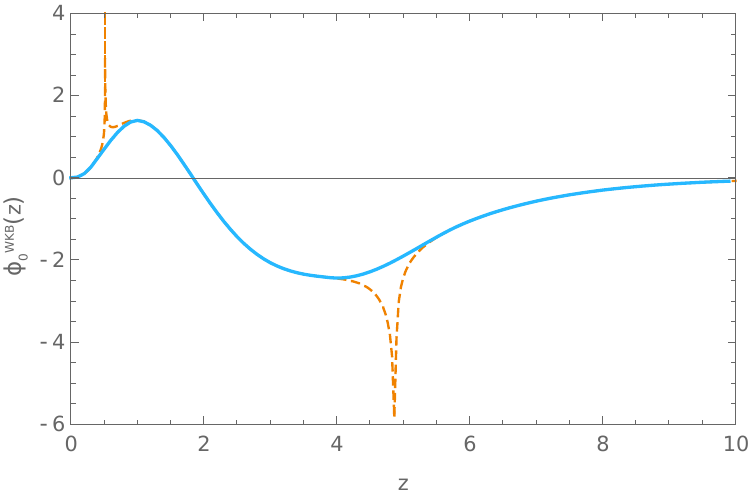}\\~\\
\includegraphics[width=7cm]{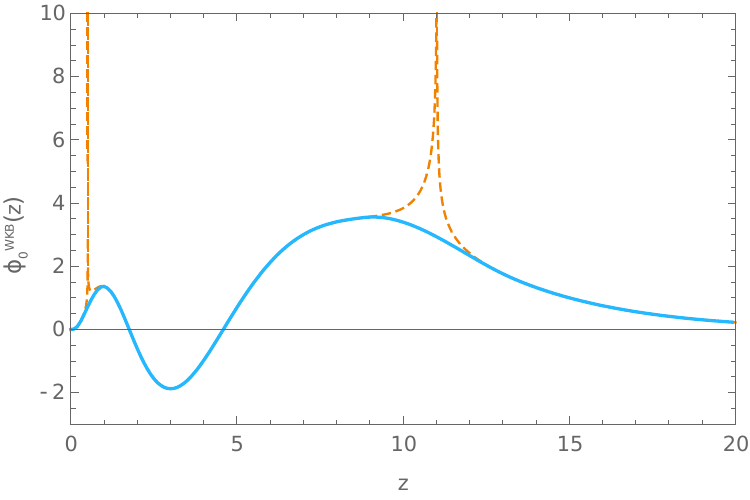}\quad
\includegraphics[width=7cm]{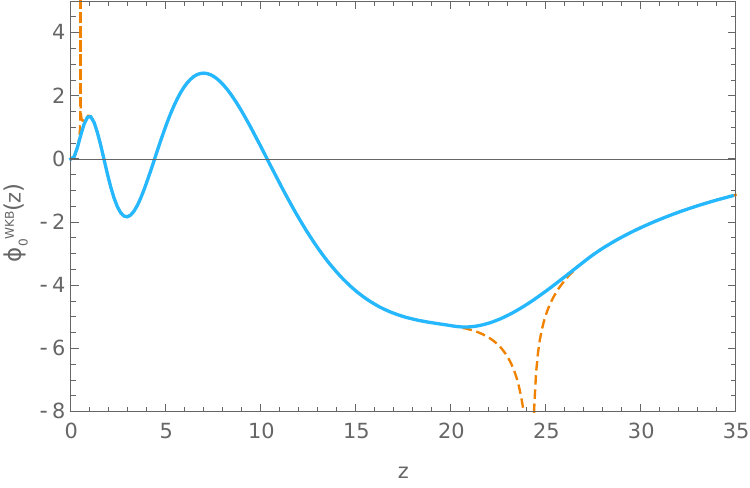}\\~\\
\includegraphics[width=7cm]{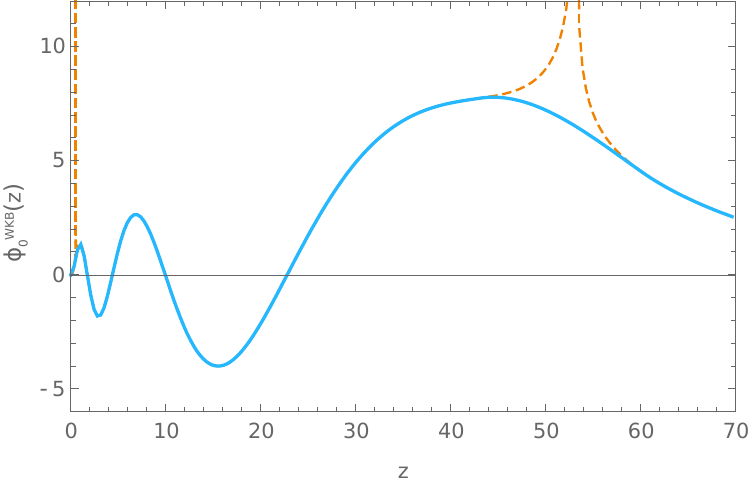}\quad
\includegraphics[width=7cm]{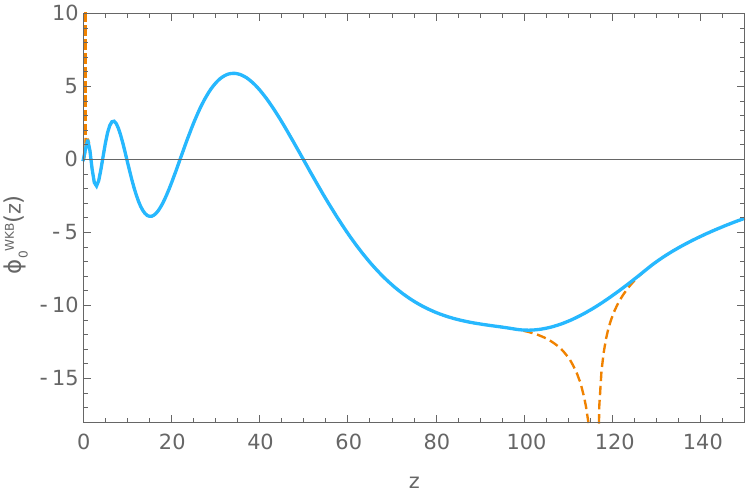}\\~\\
\includegraphics[width=7.2cm]{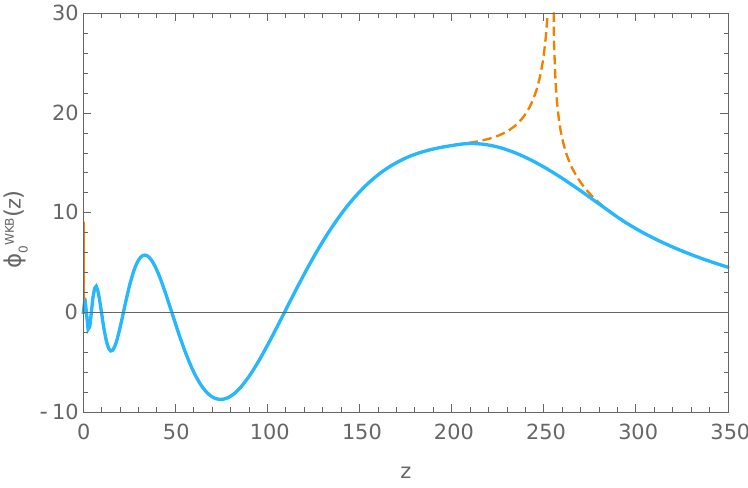}\quad
\includegraphics[width=7.3cm]{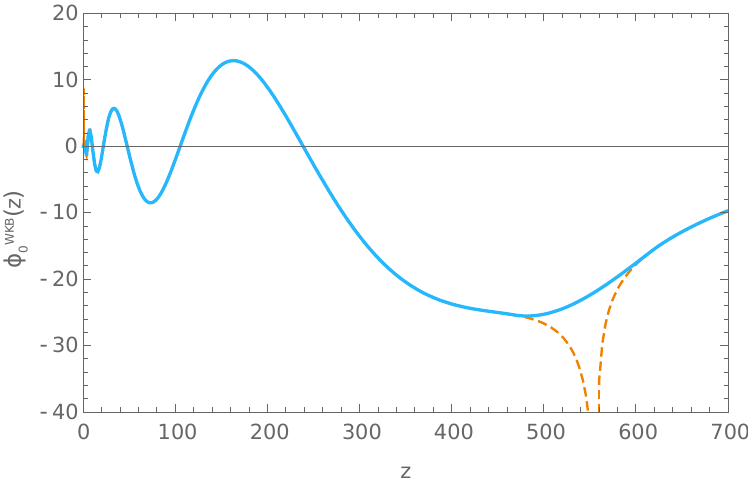}
\caption{\label{fig:WKB.electron.star.wavefunctions}
Profiles of the function $\phi_m^{\sf WKB}(z)$ are shown for the values $m\in\{0,\dots,7\}$ respectively. We see the smoothly interpolated solutions (light blue curves), and purely the WKB solutions (orange curves) that in the vicinity of the turning points were replaced by a quartic polynomial. 
}
\end{center}
\end{figure}

The connection matrix \eqref{eq:WKB.matrix.V.explicit} has only three factors ${\bf V}= {\bf M}\;{\bf W}(z_0)\;{\bf M^{\dagger}}$  and the relevant matrix element in equation \eqref{eq:Bohr-Sommerfeld.zero.freuency} takes the form
\bea\label{eq:WKB.electron.star,Bohr-Sommerfeld}
V_{11}&=&
2\,\cos{\left(
\int_{z_0}^{z_{1}} dz\,\sqrt{-U(z)}
\right)}=0\,,
\eea
implying that the Bohr-Sommerfeld condition \eqref{eq:Bohr-Sommerfeld.zero.freuency} reads
\be\label{eq:WKB.electron.star,Bohr-Sommerfeld.explicit}
\gamma\int_{z_0}^{z_1}dz\,\sqrt{ g(z)\,\left(  \frac{q^2\hat h(z)^2}{f(z)} -({k_F^{m}}^{\sf free})^2\, z^2 -\hat{m}^2 \right) } = \left(m+\frac{1}{2}\right)\,\pi\quad,\quad m=0,1,\dots\,.
\ee
In terms of an integer $m$, this equation determines the free Fermi momentum ${k_F^{m}}^{\sf free}$ of the $m$-th fermionic mode.  The results of these calculations are shown in Fig. \ref{fig:Fermi.momenta}
\vspace{-3mm}
Next, we can replace the obtained values of the free Fermi momenta ${k_F^{m}}^{\sf free}$ into equations \eqref{eq:WKB.waves}-\eqref{eq:WKB.matrix.shitfs.explicit} to obtain the corresponding solutions of  \eqref{eq:effective.schroedinger.bispinor.2} $\phi_{k\omega}^{(2)}=\phi^{\sf WKB}_m(z)$ needed to build the static wave-functions. In our calculations we found useful to replace the Airy functions interpolating around the turning points in \eqref{eq:WKB.regions} by a quartic polynomial.
Results are shown in Fig. \ref{fig:WKB.electron.star.wavefunctions}.
\subsubsection{Fermi velocities}
\label{app:WKB.electron.star.Fermi.velocities} 
To calculate the Fermi velocities $v_F^{ m\;{\sf free}}=\left.d\omega_m(k)/dk\right|_{k=k_F^{m\;{\sf free}}}$, we need to concentrate in low energy excitations, with momenta very near to the Fermi momentum $k=k_F^{\sf free}+dk$. This would result into small frequencies $\hat\omega=0+d\hat\omega$. As discussed previously, for any non-zero frequency the potential has an odd number of turning points. As our frequency is very small, along with the two turning points of the zero frequency case $z_0, z_1$, only a third one $z_2$ appears. As $d\hat\omega$ goes to zero, the third turning point $z_2$ moves to infinity.
 
In this case the resulting connection matrix $\bf V$ in \eqref{eq:WKB.matrix.V.explicit} has the five factors 
$\quad{\bf V}= {\bf M^\dagger}\;{\bf W}(z_1)\;{\bf M}\;{\bf W}(z_0)\;{\bf M^\dagger}$, 
and \eqref{eq:Bohr-Sommerfeld.finite.freuency} reads
\be\label{fig:WKB.electron.star.Bohr-Sommerfeld.velocities}
\tan\left(\,\int_{z_0}^{z_1}dz\,\sqrt{-U(z)}-\frac{\pi}{2}\right) = -\frac{i}{4}\,
e^{-2\,\int_{z_1}^{z_2(\hat\omega)}dz\,\sqrt{U(z)}}\,.
\ee
The solution for the dispersion relation is necessarily complex, defining a problem of quasi-normal modes. 
Since the integral in the exponent of the right hand side of \eqref{fig:WKB.electron.star.Bohr-Sommerfeld.velocities} diverges as $d\hat\omega$ goes to zero, the right hand side is very small. This implies that we can write \eqref{fig:WKB.electron.star.Bohr-Sommerfeld.velocities} as
\be\label{fig:WKB.electron.star.Bohr-Sommerfeld.velocities.simplified}
\int_{z_0}^{z_1}dz\,\sqrt{-U(z)}=\left(m+\frac{1}{2}
\right)\,\pi \,.
\ee
To the first order in $d\hat k$ and $d\hat \omega$, we get
\be\label{fig:WKB.electron.star.Bohr-Sommerfeld.velocities.expanded}
 \hat k_F^{\sf free} \,d\hat k\,\int_{z_0}^{z_1}dz\,\frac{g(z)z^2}{\sqrt{-U(z)}}-
 d\hat \omega\,\int_{z_0}^{z_1}dz\,\frac{qh(z)g(z)}{f(z)\sqrt{-U(z)}}
=0 \,,
\ee
implying for  the Fermi velocity
\be\label{fig:WKB.electron.star.velocities.final}
v_F^{\sf free} =  \hat k_F^{\sf free}\,\frac{\int_{z_0}^{z_1}dz\,
\frac{z^2\,g(z)}{\sqrt{-U(z)}}}{\int_{z_0}^{z_1}dz\,
\frac{g(z)\,h(z)}{f(z)\,\sqrt{-U(z)}}}\,,
\ee 
where the potential and the turning points are evaluated at zero frequency and $\hat{k}=\hat{k}_F$.

\bigskip

In is worth to mention that a possible obstacle to get straight the dispersion relation at leading order from \eqref{fig:WKB.electron.star.Bohr-Sommerfeld.velocities.simplified} arises from the non-analyticity of the complex right hand side of \eqref{fig:WKB.electron.star.Bohr-Sommerfeld.velocities}. However, as shown in \cite{Hartnoll:2011dm}, it only affects (in fact, determines!) the imaginary part of the dispersion relation, while the real part is given as usual by the analytical left hand side.
\newpage
\bibliographystyle{JHEP}

\providecommand{\href}[2]{#2}\begingroup\raggedright

\endgroup

\end{document}